\documentclass[preprint,12pt]{elsarticle}

\usepackage{amssymb}
\usepackage{amsmath}

\journal{Nuclear Physics A}

\begin{document}

\begin{frontmatter}



\title{Bohr-Mottelson Hamiltonian with octic potential applied to the $^{106-116}$Cd isotopes}


\author[a]{P. Buganu} 
 \ead{buganu@theory.nipne.ro}
\affiliation[a]{organization={Department of Theoretical Physics, "Horia Hulubei"$-$National Institute for R\&D in Physics and Nuclear Engineering},
            addressline={Reactorului 30},
            city={Bucharest-Magurele},
            postcode={RO-077125},
            country={Romania}}
\author[a,b]{R. Budaca} 
\affiliation[b]{organization={Academy of Romanian Scientists},
            addressline={Splaiul Independentei 54},
            city={Bucharest},
            postcode={P. O. 050094},
            country={Romania}}

\begin{abstract}
The Bohr-Mottelson Hamiltonian, with an octic potential in the $\beta$ deformation variable, is numerically solved  for a $\gamma$-unstable symmetry of the nuclear system. The analytical structure of the model allows the description of multiple phenomena of great interest for the nuclear structure such as ground-state shape phase transitions and their critical points, dynamical shape phase transitions, shape coexistence with and without mixing, anomalous in-band $E2$ transitions, large $E2$ intra-band transitions and large monopole transition between the first excited $0^+$ state and the ground state, respectively. As a first application of the present model is selected the $^{106-116}$Cd isotopic chain known in literature to manifest shape phase transition, respectively shape coexistence and mixing.
\end{abstract}



\begin{keyword}
Bohr-Mottelson Model, Octic potential, Shape phase transitions, Shape coexistence and mixing
\end{keyword}

\end{frontmatter}



\section{Introduction}
The Bohr-Mottelson model \cite{Bohr1,Bohr2} is probably the most suitable to describe nuclear phenomena related to the shape such as quantum shape phase transitions (QSPTs) and critical points (CPs) \cite{Fortunato1,Casten,Cejnar,Buganu1,Fortunato2} or shape coexistence and mixing \cite{Morinaga,Heyde,Martinou,Garrett,Bonatsos,Bona} because within this model the properties of the collective states emerge naturally from vibrations and rotations of the nuclear surface around an equilibrium shape. Thus, one has a direct correspondence between nuclear shape and structure of the corresponding states which can represent a great advantage in investigation and interpretation of the experimental data. On the other hand, the microscopic models provide fundamental explanations and consequently a deep understanding for the presence of these phenomena. Some examples are the Shell Model \cite{Mayer,Poves1}, the Interacting Boson Model \cite{Iachello,Nomura,Maya,Cheng}, Partial Dynamical Symmetry \cite{Leviatan}, Proxy-SU(3) \cite{Martinou,Martinou1,Martinou2}, Covariant Density Functional Theory (CDFT) \cite{Martinou2,Yang,Niksic}, ab initio approaches \cite{Zhou,Hu}, a five-dimensional collective Hamiltonian based on CDFT \cite{Majola,Matsuyanagi} and many others \cite{Bona}. Sometimes, a complementary application of both phenomenological and microscopic models to the experimental data seems to be also appropriate \cite{Mennana1,Mennana2,Benjedi,Buganu2}.

Many solutions have been given for the Bohr-Mottelson Hamiltonian since its proposal \cite{Fortunato1,Buganu1}, most of them being motivated by a better description of the critical points of the shape phase transitions, initially represented by the E(5) and X(5) symmetries \cite{Iachello1,Iachello2}. A class of these solutions, namely those associated with the polynomial potentials in the deformation variable, distinguished itself by the consistent description not only of the critical point but also of the entire shape phase transition. Notably, sextic potential can present two minima, a spherical and a deformed one. Moreover, exact solutions of the Schrodinger equation for this potential are possible, but only for a part of the eigenvalue problem through the so-called quasi-exactly solvable method \cite{Ushveridze,Budaca,Levai,Levai2,Levai3}. In order to describe other phenomena than CPs and QSPTs, a more general solution is needed, involving a numerical diagonalization procedure. The General Collective Model (GCM), which uses as a basis of diagonalization the solutions of the five-dimensional harmonic oscillator \cite{Gneuss,Gneuss2,Hess,Eisenberg}, is one such approach. An improvement in analytical and computational efficiency was later achieved through the Algebraic Collective Model (ACM) \cite{ACM1,ACM2}. A diagonalization method with a faster convergence rate was recently proposed, using as basis functions the solutions of the infinite square well potential expressed in terms of the Bessel functions of the first kind \cite{Budaca1}. Its advantages in respect to the GCM and ACM were discussed in Ref. \cite{Budaca2}. At this point, the method has been successfully applied in the description of shape coexistence and mixing, dynamical shape transition as a function of the total spin or even unusually small quadrupole electromagnetic transitions in the yrast band \cite{Mennana1,Mennana2,Benjedi,Buganu2,Budaca1,Budaca2,BudacaA,Budaca3,Buganu3}.

Recent applications to the experimental data of the Bohr-Mottelson Hamiltonian with sextic potential suggest that higher order  polynomial potentials are necessary for a better description of the experimental data and for investigating more complicated nuclear spectra. Consequently, in the present study, the Bohr-Mottelson Hamiltonian with octic potential in the $\beta$ variable is numerically diagonalized in a basis of the Bessel functions of the first kind for the special case of the $\gamma$-unstable symmetry of the nuclear system. This symmetry offers the possibility of an exact separation of variables.

The plan of the present work is the following. In Section II, the solution of the $\gamma$-unstable Bohr-Mottelson Hamiltonian with octic potential in $\beta$ variable is introduced in detail. Numerical results revealing possible scenarios relevant for nuclear level structure, as well as preliminary applications to the experimental data of the isotopes of Cadmium, are shown in Section III. Finally, the last Section IV is dedicated to the main conclusions of the present study, respectively to underline possible new perspectives of the model development.

\section{Description of the model}
The Bohr-Mottelson Hamiltonian model, which describes the nuclear collective states in terms of vibrations and rotations of the nuclear surface, has the following expression written in the intrinsic reference frame \cite{Bohr1,Bohr2}:
\begin{eqnarray}
H&=&-\frac{\hbar^{2}}{2B}\left[\frac{1}{\beta^{4}}\frac{\partial}{\partial\beta}\beta^{4}\frac{\partial}{\partial\beta}+\frac{1}{\beta^{2}\sin(3\gamma)}\frac{\partial}{\partial\gamma}\sin(3\gamma)\frac{\partial}{\partial\gamma}\right]\nonumber\\
&&+\frac{\hbar^{2}}{8B\beta^{2}}\sum_{k=1}^{3}\frac{\hat{Q}_{k}^{2}}{\sin^{2}\left(\gamma-\frac{2\pi}{3}k\right)}+V(\beta,\gamma),
\label{Hamiltonian}
\end{eqnarray}
where $\hbar$ and $B$ are the reduced Plank constant and the mass parameter, $\beta$ and $\gamma$ are the intrinsic polar coordinates describing the shape of the nucleus, while $\hat{Q}_{k}$ are the angular momentum projections which depend on the three Euler angles $\theta_{1}$, $\theta_{2}$ and $\theta_{3}$. Thus, one has a five-dimensional Hamiltonian containing kinetic terms in both $\beta$ and $\gamma$ coordinates (vibrations), a rotational term (rotations) and an energy potential. Most often, the energy potential is selected as a sum between a potential in $\beta$ and other in $\gamma$ \cite{Fortunato1,Buganu1}:
\begin{equation}
V(\beta,\gamma)=V_{1}(\beta)+\frac{1}{\beta^{2}}V_{2}(\gamma).
\end{equation}
This form of the potential allows an exact separation of the $\beta$ from the $\gamma$ and the three Euler angles, the latter still remaining coupled due to the rotational term:
\begin{equation}
\left[-\frac{1}{\beta^{4}}\frac{\partial}{\partial\beta}\beta^{4}\frac{\partial}{\partial\beta}+v_{1}(\beta)-\varepsilon+\frac{\omega}{\beta^{2}}\right]F(\beta)=0,
\label{eqbeta}
\end{equation}
\begin{eqnarray}
&&\left[-\frac{1}{\sin(3\gamma)}\frac{\partial}{\partial\gamma}\sin(3\gamma)\frac{\partial}{\partial\gamma}+\sum_{k=1}^{3}\frac{\hat{Q}_{k}^{2}}{4\sin^{2}\left(\gamma-\frac{2\pi}{3}k\right)}+v_{2}(\gamma)\right]\nonumber\\
&&\times\psi(\gamma,\theta_{i})=\omega\psi(\gamma,\theta_{i}),\;i=1,2,3.
\label{eqgamma}
\end{eqnarray}
This separation of the variables has been achieved by choosing the wave function $\Psi(\beta,\gamma,\theta_{i})=F(\beta)\psi(\gamma,\theta_{i})$, respectively by adopting the notations for the separation constant $\omega$, the reduced total energy $\varepsilon=\frac{2B}{\hbar^{2}}E$ and potentials $v_{1}(\beta)=\frac{2B}{\hbar^{2}}V_{1}(\beta)$ and $v_{2}(\gamma)=\frac{2B}{\hbar^{2}}V_{2}(\gamma)$.
Further, the solvability of the equations (\ref{eqbeta}) and (\ref{eqgamma}) is related to the treatment of the rotational term, but also to the type of potentials in $\beta$ and $\gamma$ variables. These aspects will be further addressed accordingly.

\subsection{Solution for the equation in the $\beta$ variable}
In the present study, Eq. (\ref{eqbeta}) is numerically solved for an octic potential (OP) of the form:
\begin{equation}
v_{1}(\beta)=a_{1}\beta^{2}+a_{2}\beta^{4}+a_{3}\beta^{6}+a_{4}\beta^{8},
\label{OP}
\end{equation}
where $a_{1}$, $a_{2}$, $a_{3}$ and $a_{4}$ are free parameters. The method is the same involved in previous works for the sextic potential \cite{Budaca1,Budaca2,Budaca3}. Thus, as a basis of numerical diagonalization are used the solutions for the infinite square well potential (ISWP) \cite{Bes,Iachello1,Iachello2,Bonatsos1,Bonatsos2,Bonatsos3}:
\begin{eqnarray}
	v(\beta)=\bigg \{
	  \begin{matrix}
	  0, & \beta\leq\beta_{w}, \\
	  \infty, & \beta>\beta_{w}, \\
	  \end{matrix}
\label{iswp}
\end{eqnarray}
where $\beta_{w}$ is the point where the original potential can be safely approximated with infinity. Eq. (\ref{eqbeta}) with ISWP (\ref{iswp}), after expanding the kinetic term in $\beta$, is written as:
\begin{equation}
\left(-\frac{d^{2}}{d\beta^{2}}-\frac{4}{\beta}\frac{d}{d\beta}-\varepsilon_{w}+\frac{\omega}{\beta^{2}}\right)f(\beta)=0,
\end{equation}
with $\varepsilon_{w}$ being the eigenvalue of the $\beta$ equation for the ISWP. This equation is further reduced to a Bessel form by considering the change of function $f(\beta)=\beta^{-\frac{3}{3}}J(\beta)$ and variable $z=\beta k$, respectively by adopting the notations $\varepsilon_{w}=k^{2}$ and $\nu=\sqrt{\omega+\frac{9}{4}}$:
\begin{equation}
\left[\frac{d^{2}}{dz^{2}}+\frac{1}{z}\frac{d}{dz}+\left(1-\frac{\nu^{2}}{z^{2}}\right)\right]J(z)=0.
\end{equation}
From the boundary condition $J(\beta_{w})=0$, one gets the eingenvalues and the wavefunctions for the ISWP:
\begin{equation}
\varepsilon_{w;n,\nu}=\frac{z_{n,\nu}^{2}}{\beta_{w}^{2}},\,f_{n,\nu}(\beta)=\frac{\sqrt{2}}{\beta_{w}}\frac{\beta^{-\frac{3}{2}}J_{n,\nu}\left(\frac{z_{n,\nu}}{\beta_{w}}\beta\right)}{J_{\nu+1}(z_{n,\nu})},
\end{equation}
where $z_{n,\nu}$ is the $n^{th}$ zero of the Bessel function $\tilde{J}_{n,\nu}(z)$ of index $\nu$. The wave function for the octic potential is written as a linear combination of the solutions for the ISWP \cite{Budaca1,Taseli}:
\begin{equation}
F_{\xi,\nu}(\beta)=\sum_{n=1}^{n_{Max}}A_{n}^{\xi}f_{n,\nu}(\beta),
\label{totwav}
\end{equation}
where $n_{Max}$ gives the dimension of the basis, $\xi=1,2,...$ index the diagonalization solutions and it is related to the $\beta$ vibration quantum number $(n_{\beta}=\xi-1)$, while $A_{n}^{\xi}$ are the eigenvector components. Further, by substituting the wave function (\ref{totwav}) in Eq. (\ref{eqbeta}) for the OP (\ref{OP}), one gets the general matrix element of the total Hamiltonian \cite{Budaca1,Budaca2,Budaca3,Buganu3}:
\begin{equation}
H_{mn}=\left(\frac{z_{n,\nu}}{\beta_{w}}\right)^{2}\delta_{mn}+\sum_{i=1}^{4}a_{i}\beta_{w}^{2\cdot i}I_{mn}^{(\nu,i)}.
\label{matrixelements}
\end{equation}
Here, by $I_{mn}^{(\nu,i)}$ are denoted the matrix elements for the potential terms:
\begin{equation}
I_{mn}^{(\nu,i)}=\frac{2}{J_{\nu+1}(z_{m,\nu})J_{\nu+1}(z_{n,\nu})}\int_{0}^{1}J_{\nu}(z_{m,\nu}x)J_{\nu}(z_{n,\nu}x)x^{2i+1}dx,
\end{equation}
where the change of variable $x=\beta/\beta_{w}$ was involved such that the matrix elements to be independent on $\beta_{w}$ and consequently on any parameter. This form offers an advantage in the numerical calculations because the matrix elements in $x$ can be calculated only once and than used for different values of the free parameters. From this point, in order to find the matrix elements (\ref{matrixelements}) concretely, one needs to find out the expression for $\nu$. The index $\nu$ of the Bessel function depends on the separation constant $\omega$, which in turn can have different expressions depending on the treatment of the equation in the $\gamma$ variable.

\subsection{Solution of the $\gamma$ equation in the $\gamma$-unstable case}
In the $v_{2}(\gamma)=0$ case, the non-axial $\gamma$-deformation is not stable, oscillating between prolate ($\gamma=0^{o}$) and oblate ($\gamma=60^{o}$) shapes with an extended flat peak for the probability density distribution of deformation in $\gamma$ centered around the triaxial shape value $\gamma=30^{o}$. The potential in this situation becomes $\gamma$-independent, while the separation constant,
\begin{equation}
\omega=\tau(\tau+3),\;\tau=0,1,2,...
\end{equation}
corresponds to the Casimir operator of the SO(5) group, where $\tau$ is the seniority quantum number \cite{Bohr1,Bes,Wilets}. Having the expression for $\omega$, the index $\nu$ of the wave function (\ref{totwav}) can be determined as
\begin{equation}
\nu=\sqrt{\tau(\tau+3)+\frac{9}{4}}=\tau+\frac{3}{2},
\end{equation}
respectively, the total energy to be calculated:
\begin{equation}
E_{\xi,\tau,L}=\frac{\hbar^{2}}{2B}\left[\varepsilon_{\xi,\tau}(a_{1},a_{2},a_{3},a_{4})+bL(L+1)\right].
\label{eunstable}
\end{equation}
The last term, $L(L+1)$, is the eigenvalue of the $SO(3)$ symmetry operator $\hat{L}^{2}$, which is added to the Hamiltonian to break the degeneracy of the states after $\tau$ \cite{Caprio}. For example, the lowest degenerated states for $\xi=1$ are: $(\tau=2;L=4,2)$, $(\tau=3;L=6,4,3,0)$, $(\tau=4;L=8,6,5,4,2)$ \cite{Caprio}.  Thus, this last term allows a more realistic description of the experimental data for states which are close in energy but not degenerated. The scaling parameter $\hbar^{2}/2B$ can be removed if the following energy normalization is used:
\begin{equation}
R_{\xi,\tau,L}=\frac{E_{\xi,\tau,L}-E_{1,0,0}}{E_{1,1,2}-E_{1,0,0}}
\label{norunst}
\end{equation}
The infinite barrier $\beta_w$ is not a free parameter, but depends on the coefficients $\{a_{i}\}(i=1-4)$ of the adopted octic potential. It is fixed such that to intersect the tail of the OP at a very high energy.

\subsection{Transition probabilities}

Once the energies are calculated, a special attention is dedicated to the quadrupole ($E2$) electric transitions between the states of the ground, $\beta$ and $\gamma$ bands. These are determined with the transition operator \cite{Bohr1,Bohr2,Bes}:
\begin{equation}
T_{2,\mu}^{(E2)}=\frac{3R^{2}Ze\beta_{M}}{4\pi}\beta\left[D_{\mu,2}^{(2)}(\theta_{i})\cos\gamma+\frac{1}{\sqrt{2}}\left(D_{\mu,2}^{(2)}(\theta_{i})+D_{\mu,-2}^{(2)}(\theta_{i})\right)\sin\gamma\right],
\end{equation}
expressed in terms of Wigner matrices $D(\theta_{i})$ \cite{Wigner}. The absolute units are given by the nuclear radius $R=R_{0}A^{1/3}$, charge and mass numbers $Z$ and $A$ of the considered nucleus, the elementary charge $e$, and additionally by the scaling parameter $\beta_M$ of the deformation $\beta$. In numerical calculations we consider $R_{0}=1.2$ fm.
Thus, the expression for $E2$ transition probability is written as:
\begin{eqnarray}
&&B(E2;\xi\tau L\rightarrow\xi'\tau' L')=\left(\frac{3R^{2}Ze}{4\pi}\right)^{2}\beta_{M}^{2}\times\nonumber\\
&&(\tau'\alpha'L';112||\tau\alpha L)^{2}\langle\tau|||Q|||\tau'\rangle^{2}(B_{\xi\tau;\xi'\tau'})^{2},
\label{BE2}
\end{eqnarray}
where $(\tau_{1}\alpha_{1}L_{1};\tau_{2}\alpha_{2}L_{2}||\tau_{3}\alpha_{3}L_{3})$ is the $SO(5)$ Clebsch-Gordon coefficient \cite{Rowe1},
\begin{equation}
\langle\tau|||Q|||\tau'\rangle=\sqrt{\frac{\tau}{2\tau+3}}\delta_{\tau,\tau'+1}+\sqrt{\frac{\tau+3}{2\tau+3}}\delta_{\tau,\tau'-1},
\label{tauop}
\end{equation}
is the $SO(5)$ reduced matrix element of the quadrupole moment \cite{Rowe2,Rowe3},
while the integral over $\beta$ is denoted by
\begin{equation}
B_{\xi\tau;\xi'\tau'}=\langle F_{\xi,\tau}(\beta)|\beta|F_{\xi',\tau'}(\beta)\rangle.
\end{equation}
Another important quantity for the present study is the probability density distribution of deformation,
\begin{equation}
\rho_{\xi,\tau}(\beta)=[F_{\xi,\tau}(\beta)]^{2}\beta^{4}.
\label{density}
\end{equation}
Its investigation in connection with the effective potential,
\begin{equation}
V_{eff}^{(\tau)}(\beta)=\frac{\hbar^{2}}{2B}\Big[\frac{\tau(\tau+3)+2}{\beta^{2}}+a_{1}\beta^{2}+a_{2}\beta^{4}+a_{3}\beta^{6}+a_{4}\beta^{8}\Big],
\label{effpot}
\end{equation}
and the corresponding energy levels given by Eq. (\ref{eunstable}), provides a consistent multifaceted interpretation of the nuclear dynamics. Correlations between these quantities and the monopole transition $E0$ \cite{Wood},
\begin{equation}
\rho^{2}(E0;0_{2}^{+}\rightarrow0_{1}^{+})=\left(\frac{3Z}{4\pi}\right)^{2}\beta_{M}^{4}\langle F_{2,0}(\beta)|\beta^{2}|F_{1,0}(\beta)\rangle,
\label{monopole}
\end{equation}
are also sought for a multi-messenger support of the shape coexistence and mixing in nuclei. Usually, a very low energy for the first excited $0_{2}^{+}$ state and a large monopole transition between this state and the ground state $0_{1}^{+}$ is considered as a strong signature for these phenomena.

\section{Numerical results}
The octic potential in the $\beta$ variable allows a description of multiple phenomena in nuclei related to their shape. It can simulate a single spherical minimum, an approximately spherical one, less or well deformed minima, simultaneous spherical and deformed minima, a flat shape and many other interesting situations. Thus, all these cases can cover phenomena such as ground-state shape phase transitions and their critical points, shape fluctuations in the critical point, shape coexistence with and without shape mixing, spin-dependent shape evolution, unusually small inter-band quadrupole electromagnetic transitions or very large intra-band ones, large monopole transitions and many others. Some of these aspects will be discussed in the following subsection by selecting some hypothetical scenarios with possible experimental realization, while in the next subsection the isotope chain $^{106-116}$Cd is selected as a first application to the experimental data.

\subsection{Possible scenarios for the experimental data}
\begin{figure}
	\centering
		\includegraphics[width=.8\textwidth]{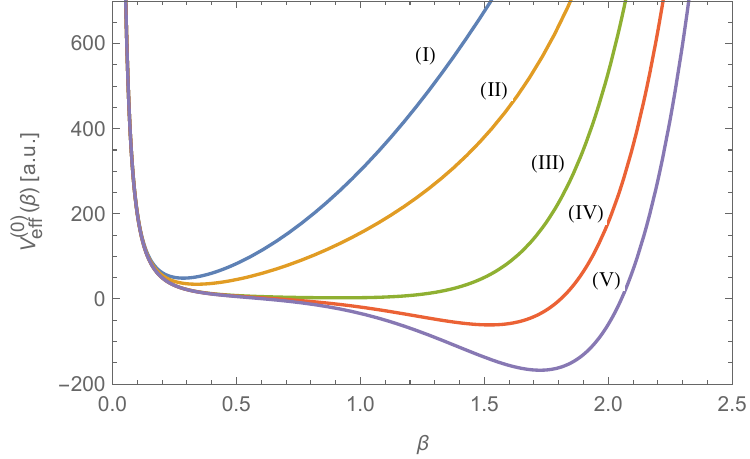} \\
	\caption{(Color online) Effective potentials (\ref{effpot}) for $\tau=0$ representing a phase transition from a spherical shape (I) to a deformed one (V), through the critical point potential (III), respectively. Curves (II) and (IV) are for intermediate shapes. The plots correspond to: $a_{1}=300,150,1,1,1$; $a_{2}=0,1,-3,-25,-40$; $a_{3}=0,1,1,1,1$, $a_{4}=0,1,2,2,2$.}
	\label{fig1}
\end{figure}

The transitional ability of the octic potential is demonstrated in Fig. \ref{fig1}, which shows the ground state effective potential given by Eq. (\ref{effpot}) for different sets of parameters. The parameters are adjusted such that to simulate a phase transition from a spherical nucleus (I) to a well-deformed one (V), crossing the critical point (III), where the potential has a flat shape. The intermediate potentials (II) and (IV) showcase the smoothness of the transition between the two shape phases. This is a second-order shape phase transition \cite{Iachello1} according to calculations made with the coherent state functions applied for the Interacting Boson Hamiltonian \cite{Iachello,Ginocchio,Dieperink}.

As not all $\{a_{i}\}(i=1-4)$ parameters are necessary for the description of a unique spectral structure, it is more convenient to perform further numerical applications within a restricted set of independent parameters. In previous studies \cite{Budaca1,Budaca2,Budaca3}, it was shown that $a_{1}$ can be extracted as a scaling parameter by adopting a change of variable $\beta=a_{1}^{-\frac{1}{4}}\tilde{\beta}$. Thus, the total energy can be expressed as
\begin{equation}
E_{\xi,\tau,L}=\frac{\hbar^{2}}{2\tilde{B}}\left[\varepsilon_{\xi,\tau}(1,b_{2},b_{3},b_{4})+\tilde{b}L(L+1)\right].
\label{eunstable1}
\end{equation}
where
\begin{equation}
\tilde{B}=\frac{B}{\sqrt{a_{1}}},\,b_{2}=a_{1}^{-\frac{3}{2}}a_{2},\,b_{3}=a_{1}^{-2}a_{3},\,b_{4}=a_{1}^{-\frac{5}{2}}a_{4},\,\tilde{b}=\frac{b}{\sqrt{a_{1}}}.
\end{equation}
The normalized energies (\ref{norunst}) will then be uniquely determined just by the three independent parameters $b_{2}$, $b_{3}$, $b_{4}$, plus an additional one $\tilde{b}$. The transformation of the wave-function and of the relevant matrix elements is straightforward.

\begin{figure}
	\centering
		\includegraphics[width=.63\textwidth]{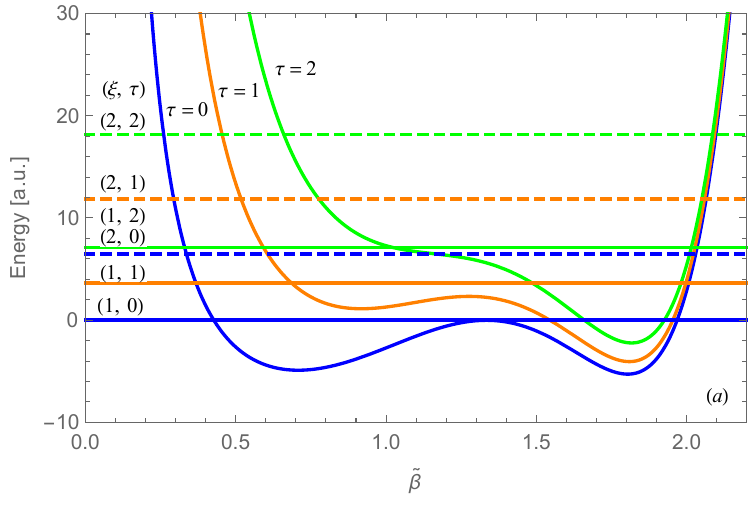} \\
        \includegraphics[width=.63\textwidth]{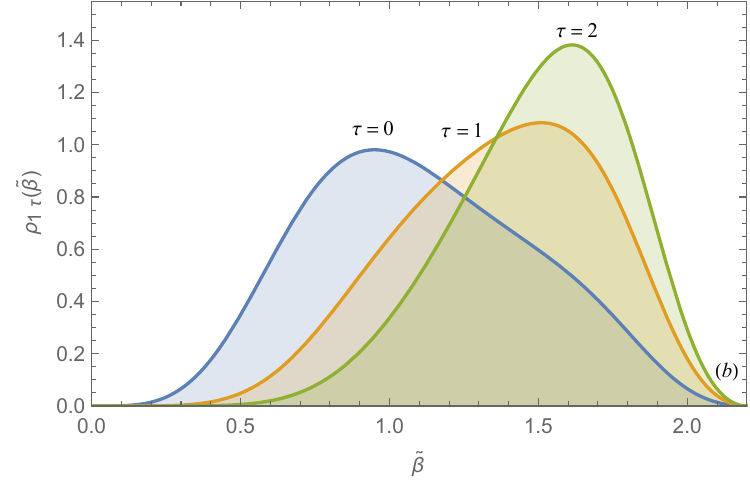} \\
        \includegraphics[width=.63\textwidth]{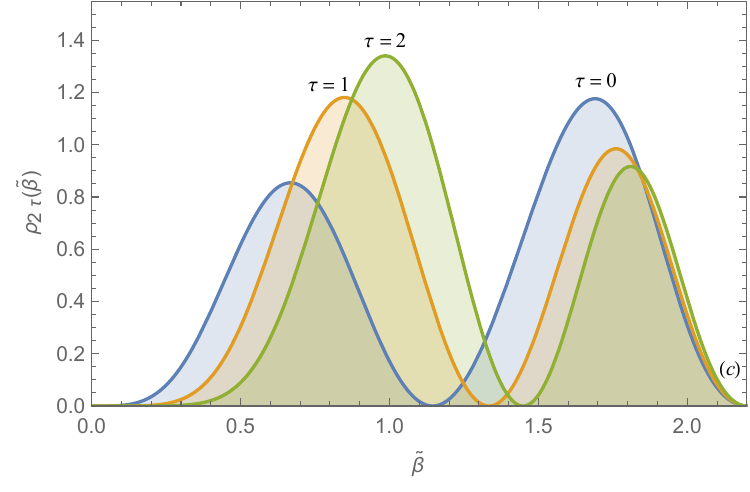} \\
	\caption{(Color online) In panel (a) are given the energy levels (\ref{eunstable1}) for the ground band $(\xi=1,\tau=0,1,2)$ and $\beta$ band $(\xi=2,\tau=0,1,2)$ with the corresponding effective potential (\ref{effpot}) as a function of the scaled $\tilde{\beta}$ variable. Additionally, in panels (b) and (c), are plotted the probability density distribution of deformation (\ref{density}) for the ground and $\beta$ band states. The results are obtained for $b_{2}=11.5$, $b_{3}=-6.72$, $b_{4}=1$ and $\tilde{\beta}_{w}=2.2$.}
	\label{fig2}
\end{figure}

In the case of a first-order shape phase transition, the critical potential has two degenerated minima, a spherical and a deformed one, separated by a very small maximum (barrier) \cite{Iachello2}. Such a transition usually commence within stable axial deformation (prolate) rather than the $\gamma$-unstable symmetry. Nevertheless, this scenario is very well illustrated in Fig. \ref{fig2}, where the lowest energy levels (\ref{eunstable}) of the ground band $(\xi=1,\tau=0,1,2)$ and $\beta$ band $(\xi=2,\tau=0,1,2)$ are plotted together with the corresponding effective potentials (\ref{effpot}) in panel (a), as well as the probability density distribution of deformation in panels (b) and (c).

The energy level of the ground state $(1,0)$ is tangent to the small barrier of the potential and the first excited state $(1,1)$ of the ground band is close to the top of the barrier, while the other states are positioned above the maximum, with a week influence from the barrier. Also, the approximately spherical minimum starts to vanish for $\tau=2$. This behavior is very well reflected in the plots for the probability density distribution of deformation from panels (b) and (c) of Figure \ref{fig2}. Both $\rho_{1,0}(\tilde{\beta})$ and $\rho_{1,1}(\tilde{\beta})$ present extended peaks with a tendency of splitting for the ground state. A well contoured peak, positioned mostly above the well-deformed minimum, is obtained for the state $(1,2)$, while the densities for the $\beta$ band states manifest a regular pattern associated with the $\beta$-vibration for $\xi=2$ (one-node). This picture is much closer to the description of the critical point of the first-order shape phase transition than that offered by the X(5) model \cite{Iachello2} or other previous models where the barrier (maximum) is neglected. For example, by disregarding this maximum and approximating the potential with an infinite square well \cite{Iachello2}, one gets an analytical solution, which up to a scaling factor is parameter free. This result represents a great tool and reference point for the interpretation of the experimental data, offering signatures for finding candidate nuclei for this critical point. Concluding the analysis of Fig. {\ref{fig2}}, the phenomenon described here is that of shape fluctuations since one has almost an extended peak for the ground state around the two degenerated minima.

\begin{figure}
	\centering
		\includegraphics[width=.65\textwidth]{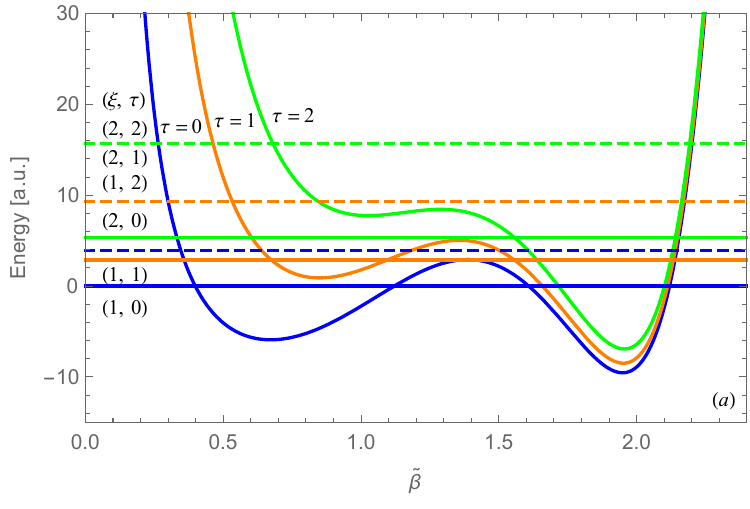} \\
        \includegraphics[width=.65\textwidth]{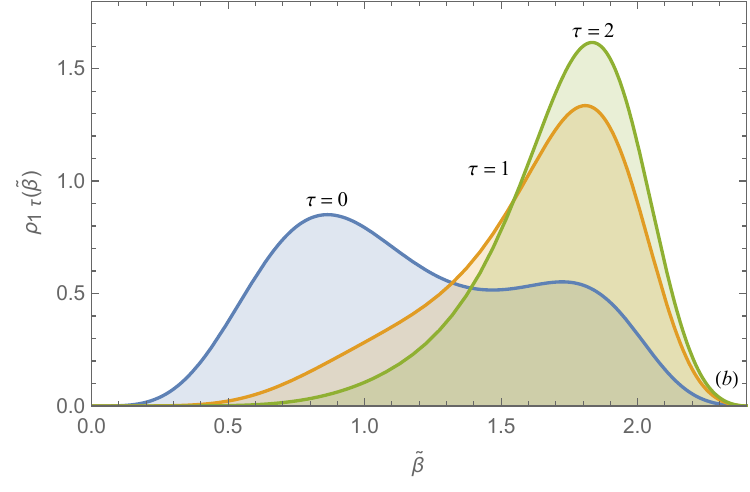} \\
        \includegraphics[width=.65\textwidth]{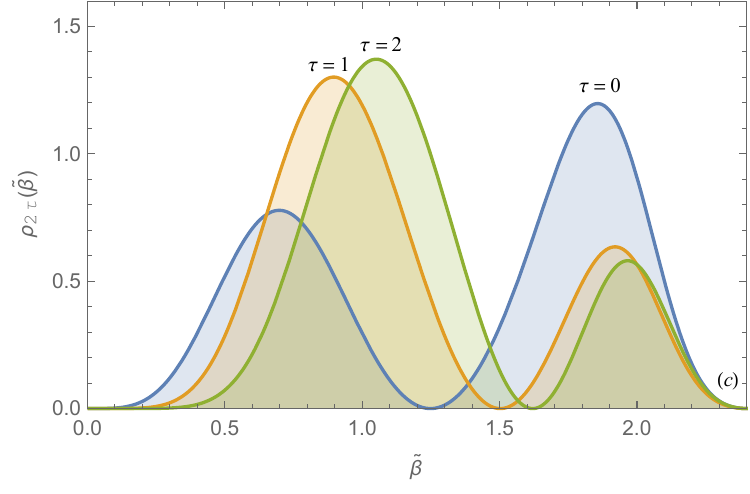} \\
	\caption{(Color online) The same as in Fig. \ref{fig2}, but for $b_{2}=14.5$, $b_{3}=-7.63$, $b_{4}=1$ and $\tilde{\beta}_{w}=2.4$.}
	\label{fig3}
\end{figure}

\begin{figure}
	\centering
		\includegraphics[width=.65\textwidth]{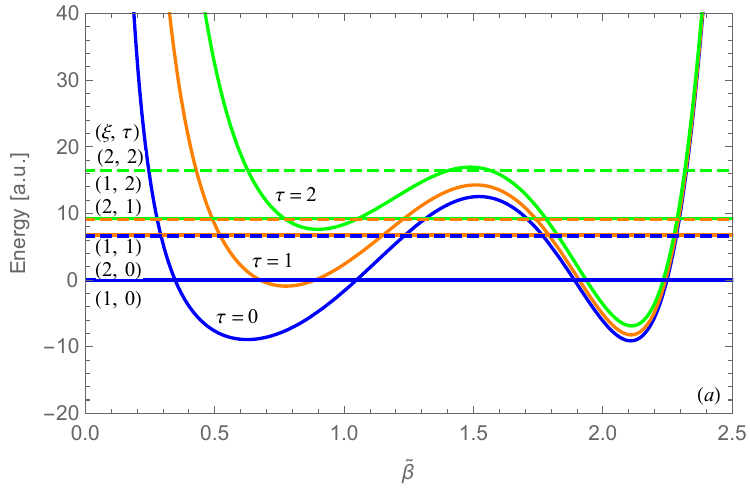} \\
        \includegraphics[width=.65\textwidth]{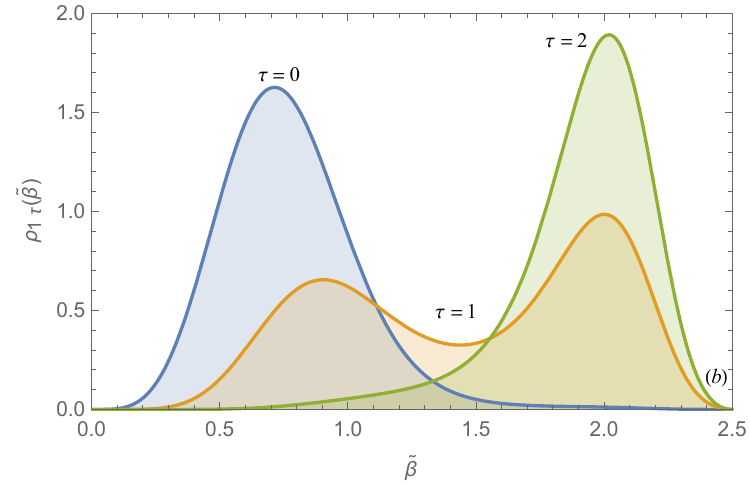} \\
        \includegraphics[width=.65\textwidth]{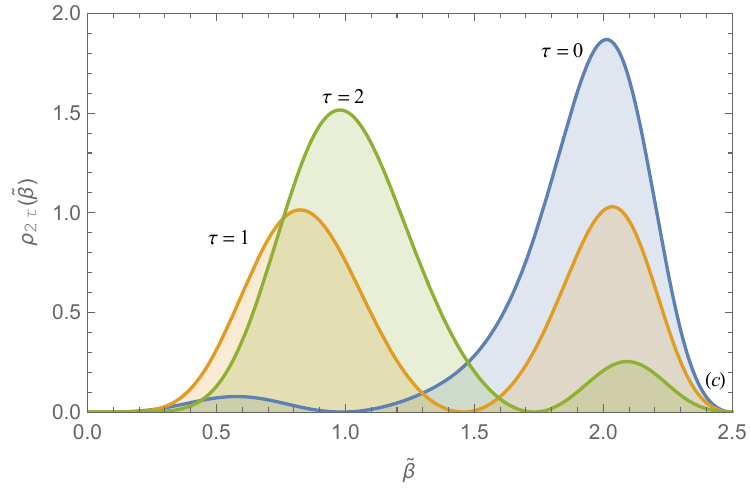} \\
	\caption{(Color online) The same as in Fig. \ref{fig2}, but for $b_{2}=20.4$, $b_{3}=-9$, $b_{4}=1$ and $\tilde{\beta}_{w}=2.5$.}
	\label{fig4}
\end{figure}

\begin{figure*}
	\centering
		\includegraphics[width=.5\textwidth]{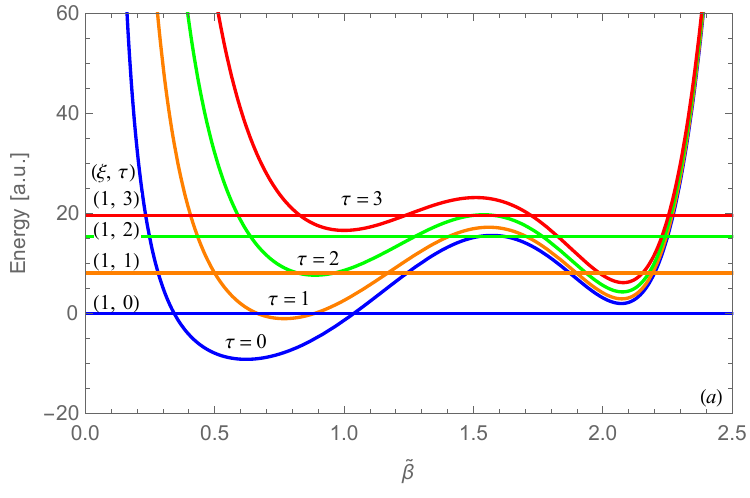}\includegraphics[width=.5\textwidth]{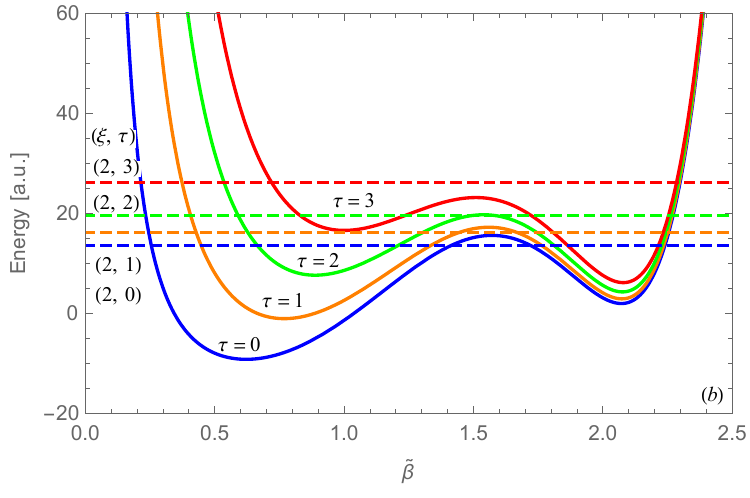} \\
        \includegraphics[width=.5\textwidth]{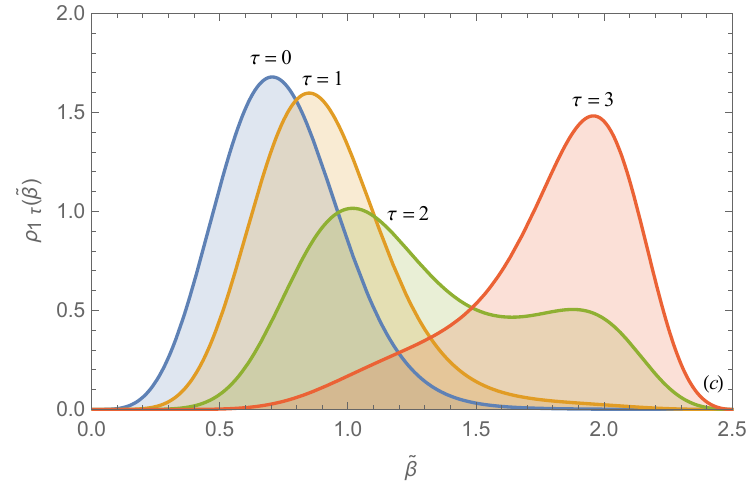}\includegraphics[width=.5\textwidth]{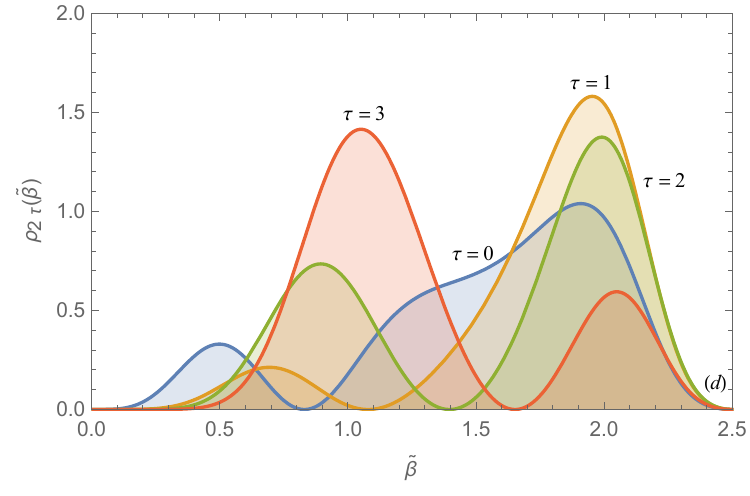}\\
	\caption{(Color online) States $(1,3)$ and $(2,3)$ are added here compared to Fig. \ref{fig2}, as well as their corresponding potentials and densities for $\tau=3$. Plots are done for $b_{2}=21$, $b_{3}=-9$, $b_{4}=1$ and $\tilde{\beta}_{w}=2.5$.}
	\label{fig5}
\end{figure*}

\begin{figure}
	\centering
		\includegraphics[width=.65\textwidth]{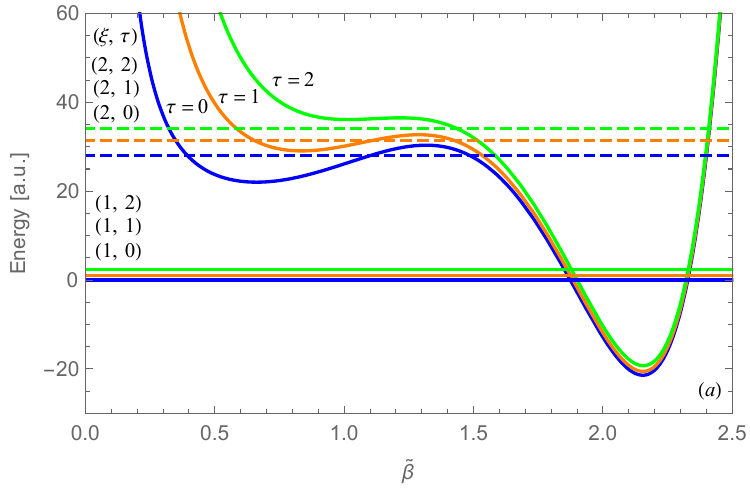} \\
        \includegraphics[width=.65\textwidth]{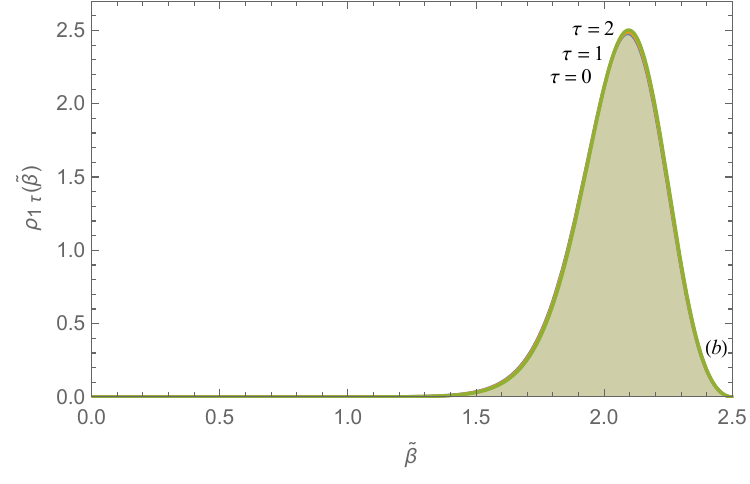} \\
        \includegraphics[width=.65\textwidth]{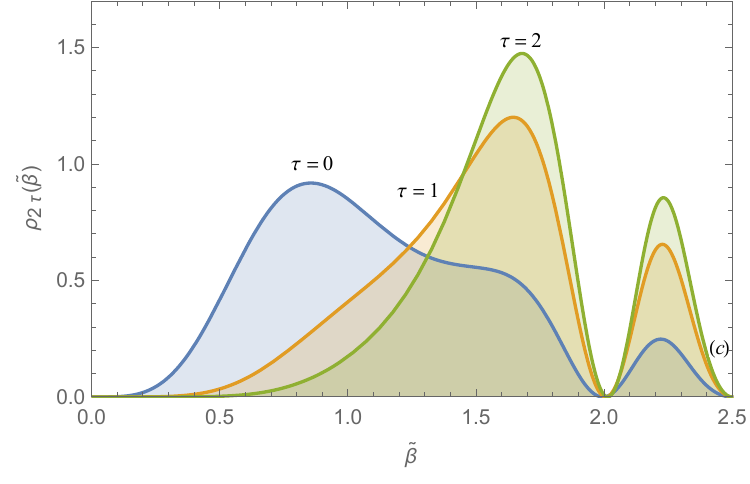} \\
	\caption{(Color online) The same as in Fig. \ref{fig2}, but for $b_{2}=16$, $b_{3}=-8.5$, $b_{4}=1$ and $\tilde{\beta_{w}}=2.5$.}
	\label{fig6}
\end{figure}

\begin{figure}
	\centering
		\includegraphics[width=.65\textwidth]{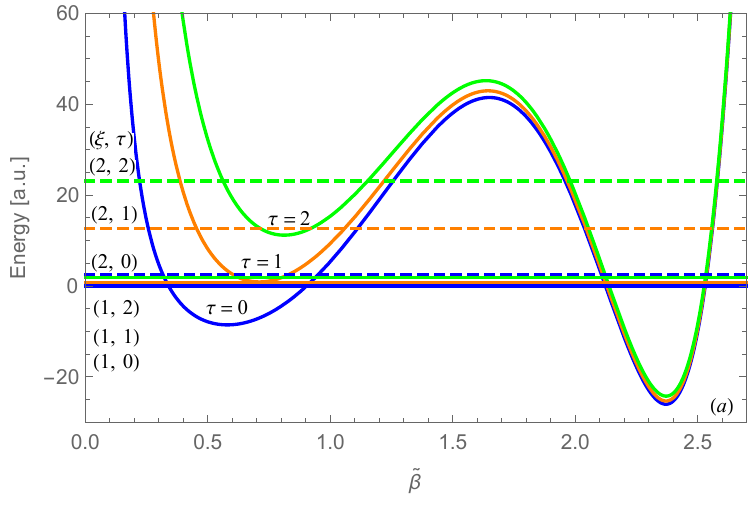} \\
        \includegraphics[width=.65\textwidth]{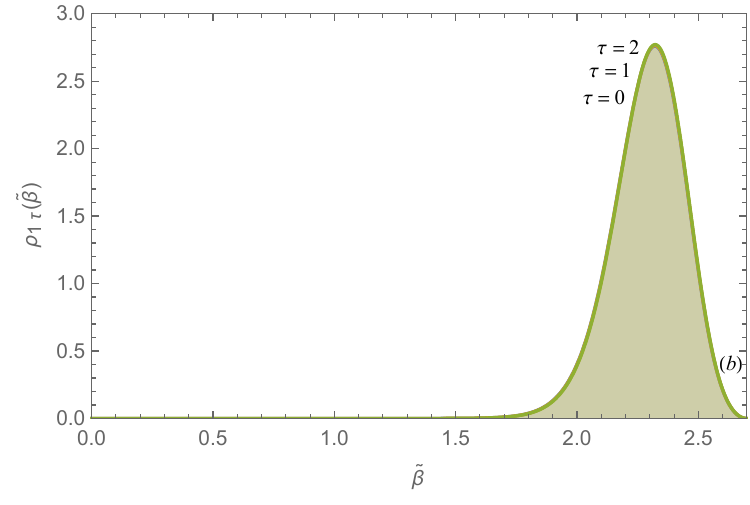} \\
        \includegraphics[width=.65\textwidth]{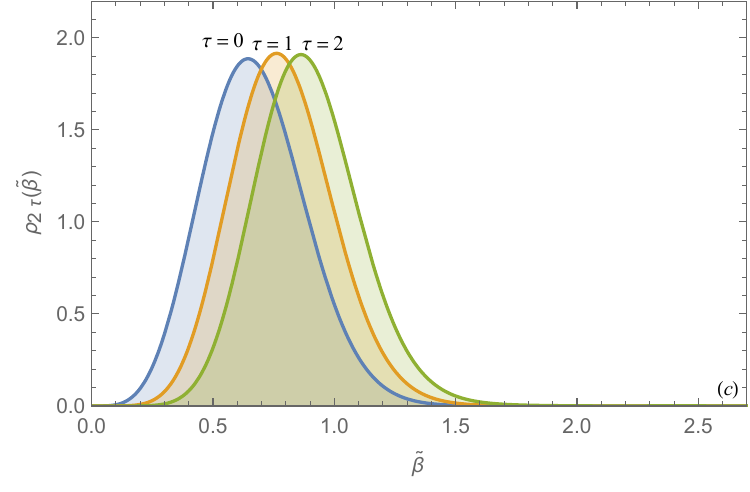} \\
	\caption{(Color online) The same as in Fig. \ref{fig2}, but for $b_{2}=30.35$, $b_{3}=-11.1$, $b_{4}=1$ and $\tilde{\beta}_{w}=2.7$.}
	\label{fig7}
\end{figure}

\begin{figure}
	\centering
		\includegraphics[width=.65\textwidth]{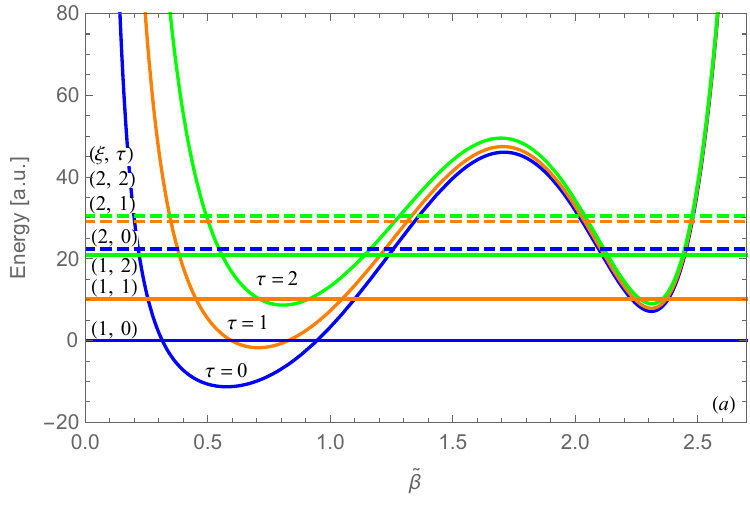} \\
        \includegraphics[width=.65\textwidth]{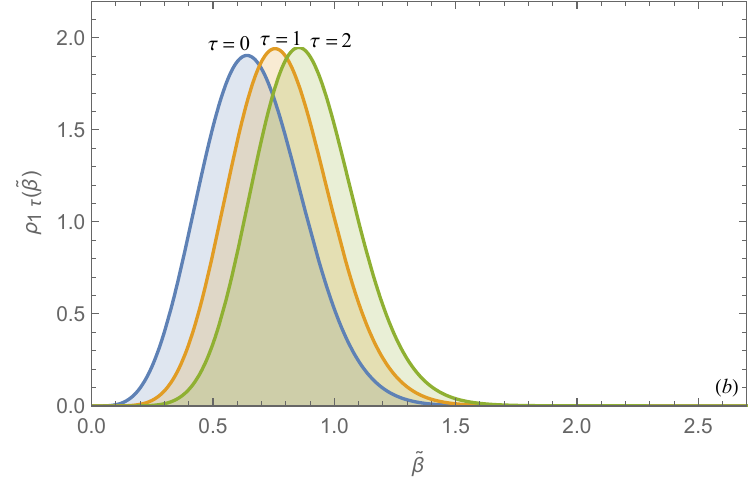} \\
        \includegraphics[width=.65\textwidth]{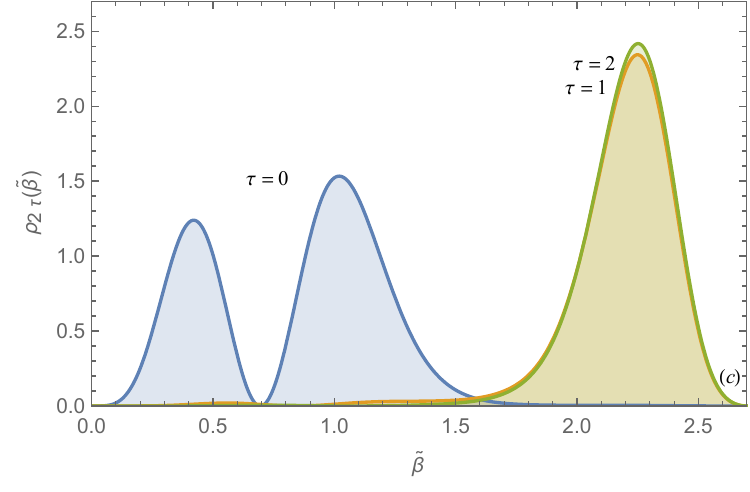} \\
	\caption{(Color online) The same as in Fig. \ref{fig2}, but for $b_{2}=31$, $b_{3}=-11$, $b_{4}=1$ and $\tilde{\beta}_{w}=2.7$.}
	\label{fig8}
\end{figure}

Now, any increase of the maximum or breaking of the degeneracy of the two minima shift the model description toward more exotic phenomena. In the following, a part of these possibilities are depicted in Figures \ref{fig3}-\ref{fig8}. The considered scenarios have the role of showing the vast ability of the model to describe distinct dynamical characteristics of the experimental data. It is expected also that pointed numerical applications will provide other interesting spectral features.

Thus, in Fig. \ref{fig3}, panel (a), the barrier is slightly increased such that the ground state $(1,0)$ falls under the barrier while the excited state $(2,0)$ comes closer to it but still above it. Also, the degeneracy of the two minima is broken in favor of the deformed minimum. This change finalizes the splitting of the peak in panel (b) for $\rho_{1,0}(\tilde{\beta})$ started in Fig. \ref{fig2} for the critical point. Moreover, in panel (c), $\rho_{1,0}(\tilde{\beta})$ presents a more dominant peak above the well-deformed minimum compared with the above excited states, which looks like in the mirror with that for $\rho_{1,0}(\tilde{\beta})$. The latter has the dominant peak above the less-deformed minimum. Thus, one can speak here about distinct shapes, on one hand, but also about a superposition between the two shapes at least for the states $(1,0)$ and $(2,0)$. The phenomenon observed is specific to shape coexistence with mixing in the ground state and first excited $0^{+}$ state. The similar situation was found in $^{74}$Ge and $^{74}$Kr nuclei \cite{Mennana1} by applying the sextic potential.

In Fig. \ref{fig4}(a), one comes back to the degeneracy of the two minima, but in turn the barrier is further increased until all the represented levels are below it. This effect induces a rotational shape evolution, from spherical to deformed shape in the ground band and viceversa in the $\beta$ band with the transition point given in each case by the state of $\tau=1$. The distinct shapes obtained for states with similar energy, as well as the emergence of two peaks this time for $(1,1)$ (no-node), symmetric peaks for $(2,1)$ and almost a single peak instead of two for $(2,0)$ (one-node) indicate the presence of both coexistence with and without mixing, respectively. Such a dynamical shape evolution was observed, for example in $^{98}$Mo \cite{Budaca2} and $^{72,74,76}$Se \cite{Budaca3} nuclei within the sextic potential formalism.

A splitting in two peaks for the probability density distribution of deformation corresponding to an excited state of the ground band can take place even for a larger angular momentum. This is achieved in Figure \ref{fig5} for a higher well-deformed minimum and corresponding barrier. Here, from panels (a) and (c), one can observe that the point of the shape transition is achieved for the state $(1,2)$ instead of $(1,1)$ as it is the case in Figure \ref{fig4}. An interesting remark is that, in panel (d), the density for the state $(2,0)$ has three peaks instead of two which usually correspond to one node of the wave function. This is associated with shape coexistence with mixing. Also in the $\beta$ band one has a shape evolution with the spin, the states for $\tau=1$ and $2$ having the dominant peak above the well-deformed minimum compared to the state $(2,3)$ which prefers the lower deformation. This case with two peaks for $(1,2)$, respectively three peaks for $(2,0)$ is very similar to the picture seen in $^{42}$Ca \cite{Benjedi}.

On the other hand, if the well-deformed minimum is significantly lower in energy than the less-deformed one, as in Figures \ref{fig6} and \ref{fig7}, the ground state remains captive in the deeper potential pocket. Some difference in behavior appear for the states characterized by $\xi=2$. In Figure \ref{fig6}, where the pocket for the less-deformed minimum of the potential is not so deep, one has three peaks for the state $(2,0)$ indicating a presence of the shape coexistence with mixing. On contrary, in Figure \ref{fig7}, also the states for $\xi=2$ are trapped above the first minimum and, more important, only a single peak shows up even if one has one node for the corresponding wave functions. Basically, the $\beta$-vibration is frozen due to the very high barrier separating the two minima limiting in this way the vibration motion. This is a good example of shape coexistence without mixing.

Compared to Figure \ref{fig7}, in Figure \ref{fig8}, the first minimum is deeper and has a very high barrier separating it from the deformed minimum. This configuration favours a reversal of the states with those of the ground band above the less-deformed minimum and those of the $\beta$ band predominantly above the well-deformed minimum. An exception represents the state $(2,0)$ which vibrates only around the first minimum for which one has a large width suitable for this motion. However, due to the very high barrier, its oscillation cannot be extended also above the second minimum. This last representation fits very well the results obtained for $^{44}$Ca in \cite{Benjedi}.

In conclusion, a part of these scenarios already have some correspondence with the experimental data through previous applications of the sextic potential for nuclei known in literature to manifest such phenomena, while many other features of the quadrupole collective states remain to be discovered by new applications of the present solution of the Bohr-Mottelson Hamiltonian with octic potential. One expects to investigate nuclei for which the sextic potential could not reproduce in totality the corresponding experimental data. Such an example will be discussed in the next subsection.

\subsection{Application to the $^{106-116}$Cd isotope chain}
A first application of the presently proposed model is considered for the $^{106-116}$Cd isotopic chain, for which previously the Bohr-Mottelson Hamiltonian with sextic potential was applied to investigate the presence of a second-order ground-state shape phase transition \cite{Lahbas}. The latter solution was based on a quasi-exactly solvable method \cite{Ushveridze} which, due to the constraints imposed on the potential parameters to get analytical solutions, cannot cover phenomena such as shape coexistence and mixing. Microscopic models were also used for the study of the Cadmium isotopes revealing the presence of these phenomena \cite{Kumar,Garrett2,Nomura2,Sharma}. It should be noted that, at the beginning and even for a long time afterward, some of these isotopes of Cadmium were given as good examples of spherical vibrator mode of motion \cite{Bohr3}. However, this assumption started to be questioned with the experimental identification of new states called intruder \cite{Bernard,Peker}. New ideas have been proposed for their interpretation, such as mixing \cite{Heyde2,Casten2}  or multiple-shape coexistence \cite{Garrett2,Wrzosek} of the two configurations, spherical vibrational and intruder ones. Other recent works have advanced more in this direction by relaxing the dominant character of harmonic vibrations on the states. This feature is instead preserved for the main low-lying states which would coexists with the $\gamma$-soft band of the intruder states \cite{Leviatan2,Gavrielov,Wang}. Actually, this picture is common to that investigated within the present model, where shape coexistence with mixing between spherical and $\gamma$-unstable representation of the states is considered. There is also a completely different approach, which proposes a reconsideration of the concept of the $\beta$-vibration, respectively of the nature of the first $0_{2}^{+}$ excited states \cite{Garrett3,Garrett4,Sharpey}. These approaches emphasize the role of the pairing correlations in the emergence of the excited states. The idea is supported by the lack of true vibrational nuclei and was discussed in connection also with the presently considered isotopes of Cd \cite{Garrett5}.

\begin{figure*}
	\centering
		\includegraphics[width=1.\textwidth]{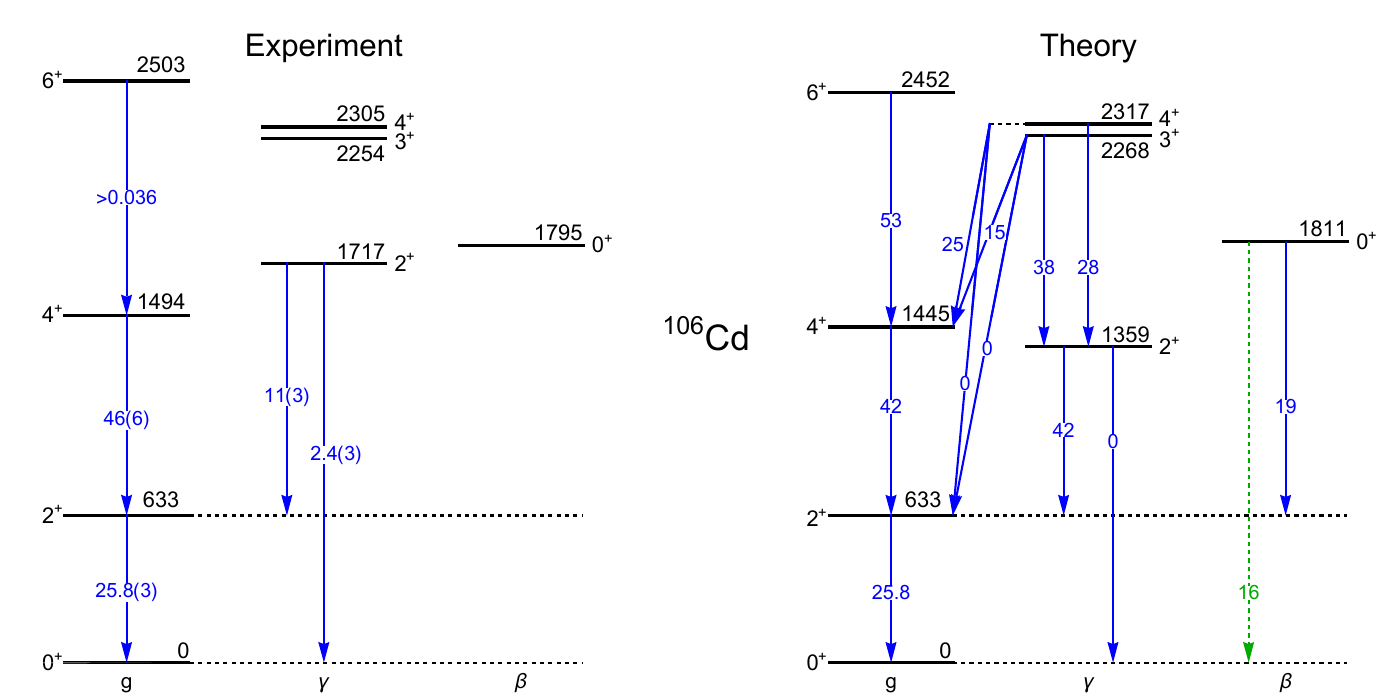} \\
	\caption{(Color online) The lowest energy levels, as well as the electromagnetic $E2$ (full arrows) and $E0$ (dashed arrow) transitions, calculated with the formulas given by Eqs. (\ref{eunstable}), (\ref{norunst}), (\ref{BE2}) and (\ref{monopole}), are compared with the experimental data of the $^{106}$Cd isotope \cite{Frenne}. The forbidden $E2$ transitions, according to the selection rules given by Eq. (\ref{tauop}), are marked with 0.}
	\label{fig9}
\end{figure*}

\begin{figure*}
	\centering
\includegraphics[width=.65\textwidth]{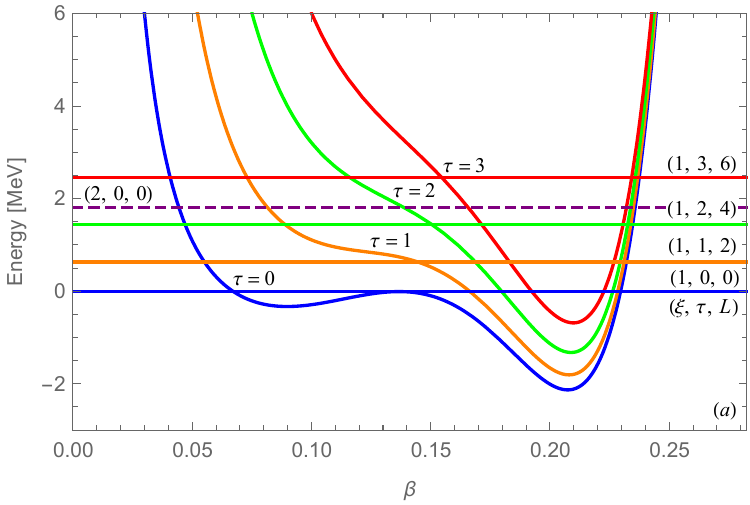}\\
\includegraphics[width=.65\textwidth]{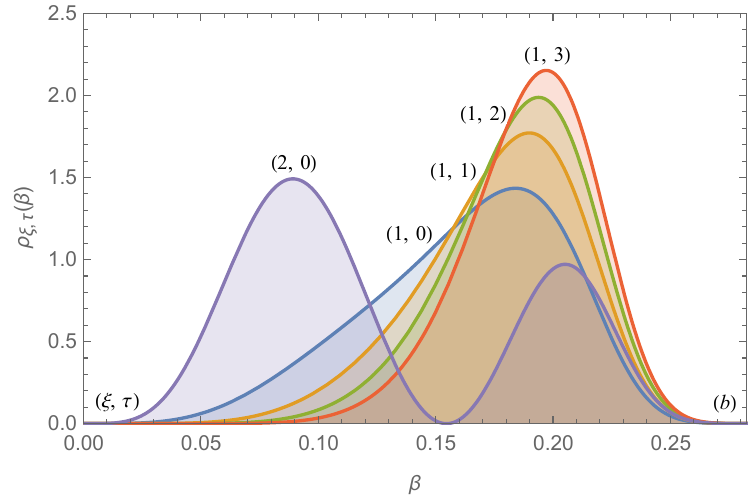}
	\caption{(Color online) Energy levels (\ref{eunstable}) in units of MeV of the ground band (full lines) and of the first excited $0^+$ state (dashed line), indexed by $(\xi,\tau,L)$, are plotted in panel (a) with the corresponding effective potentials (\ref{effpot}) as a function of $\beta=\beta_{M}\tilde{\beta}$, using the model parameters fitted for the experimental data of the $^{106}$Cd isotope \cite{Frenne}. Additionally, in panel (b), is given the probability density distribution of deformation (\ref{density}) for each state from panel (a).}
	\label{fig10}
\end{figure*}

\begin{figure*}
	\centering
		\includegraphics[width=1.\textwidth]{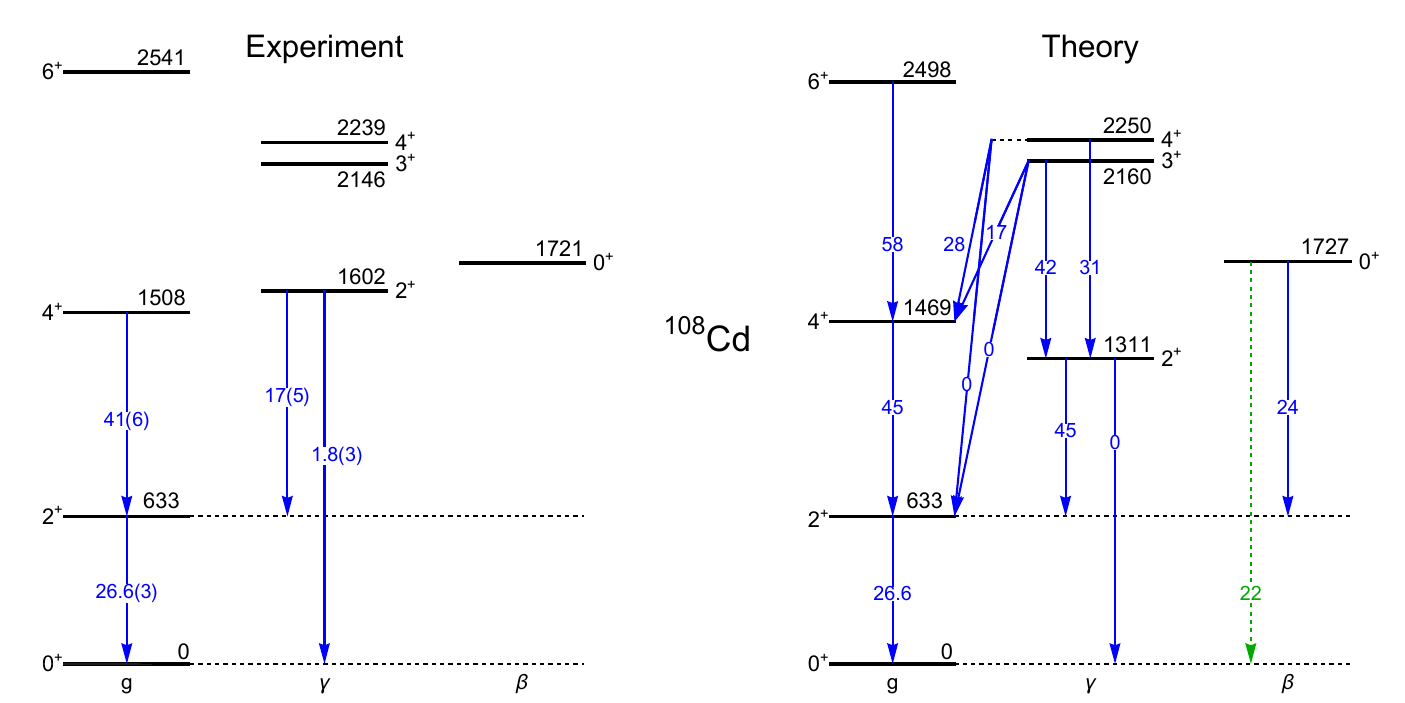} \\
	\caption{(Color online) The same as in Fig. \ref{fig9}, but for the experimental data of $^{108}$Cd \cite{Blachot1}.}
	\label{fig11}
\end{figure*}

\begin{figure*}
	\centering
\includegraphics[width=.65\textwidth]{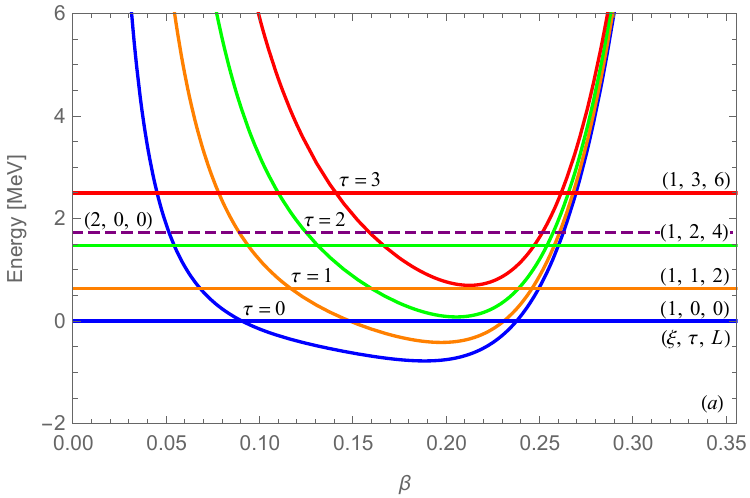}\\
\includegraphics[width=.65\textwidth]{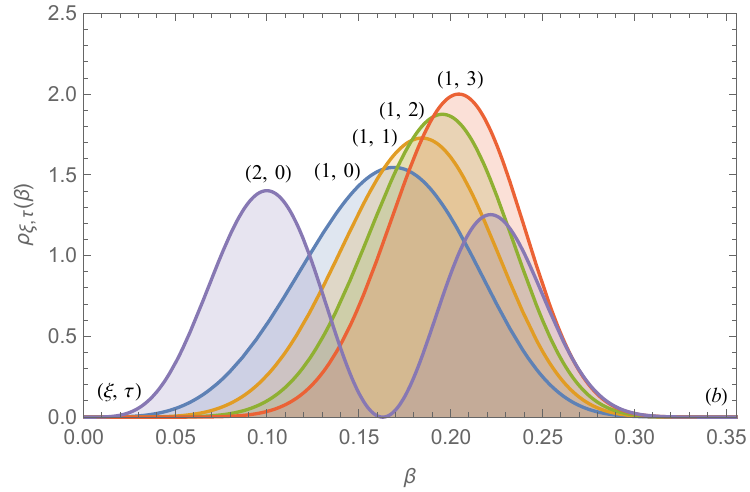}
	\caption{(Color online) The same as in Fig. \ref{fig10}, but for $^{108}$Cd \cite{Blachot1}.}
	\label{fig12}
\end{figure*}

\begin{figure*}
	\centering
		\includegraphics[width=1.\textwidth]{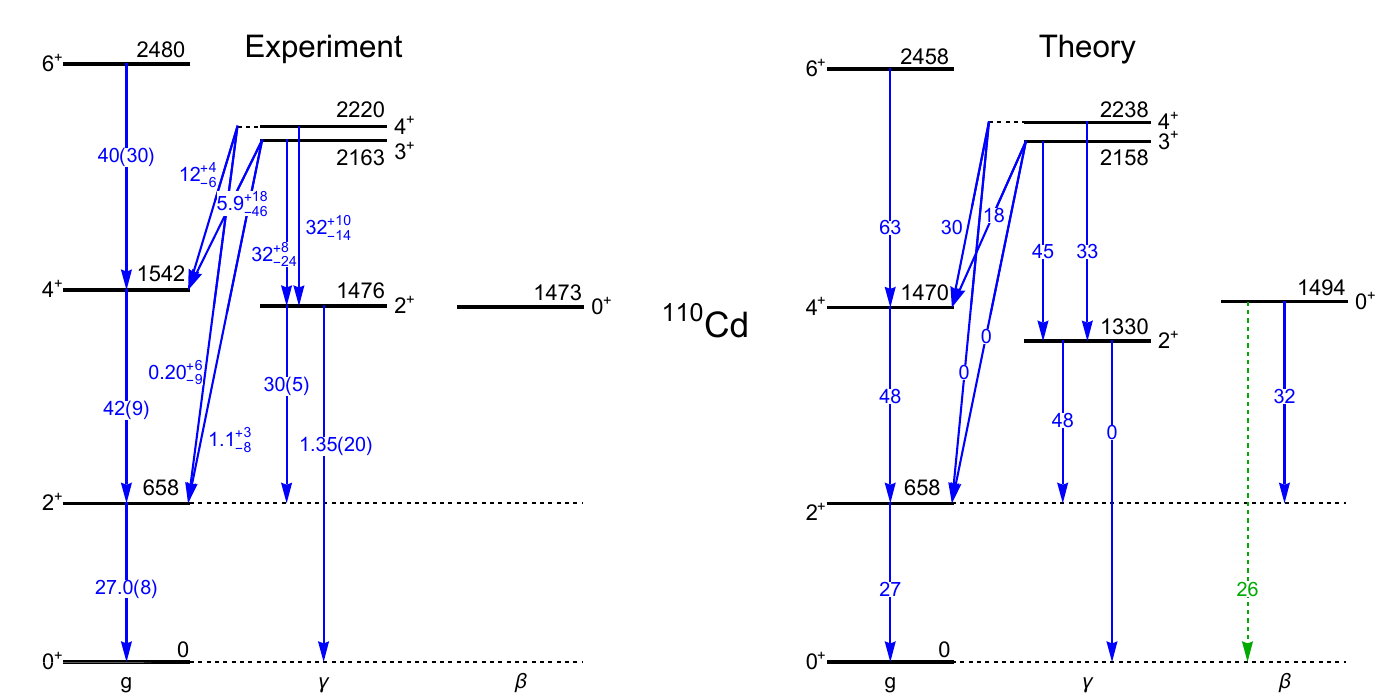} \\
	\caption{(Color online) The same as in Fig. \ref{fig9}, but for the experimental data of $^{110}$Cd \cite{Gurdal}}
	\label{fig13}
\end{figure*}

\begin{figure*}
	\centering
\includegraphics[width=.65\textwidth]{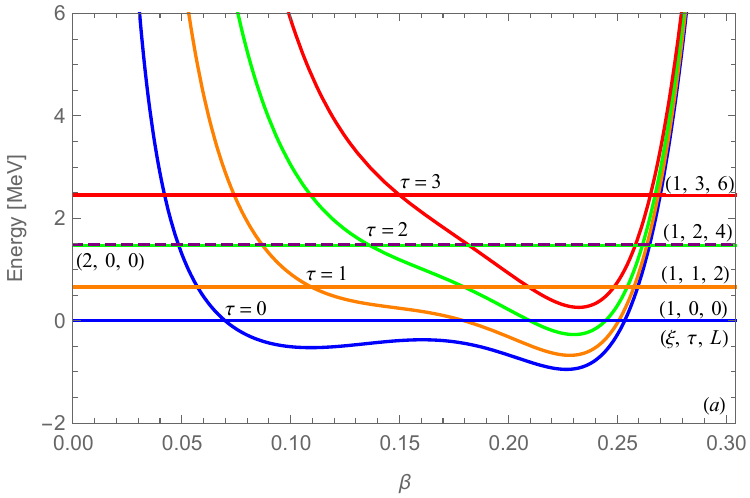}\\
\includegraphics[width=.65\textwidth]{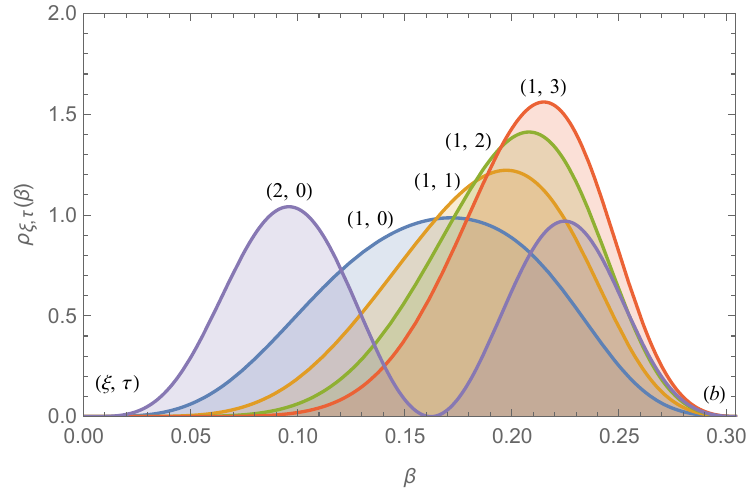}
	\caption{(Color online) The same as in Fig. \ref{fig10}, but for $^{110}$Cd \cite{Gurdal}.}
	\label{fig14}
\end{figure*}

\begin{figure*}
	\centering
		\includegraphics[width=1.\textwidth]{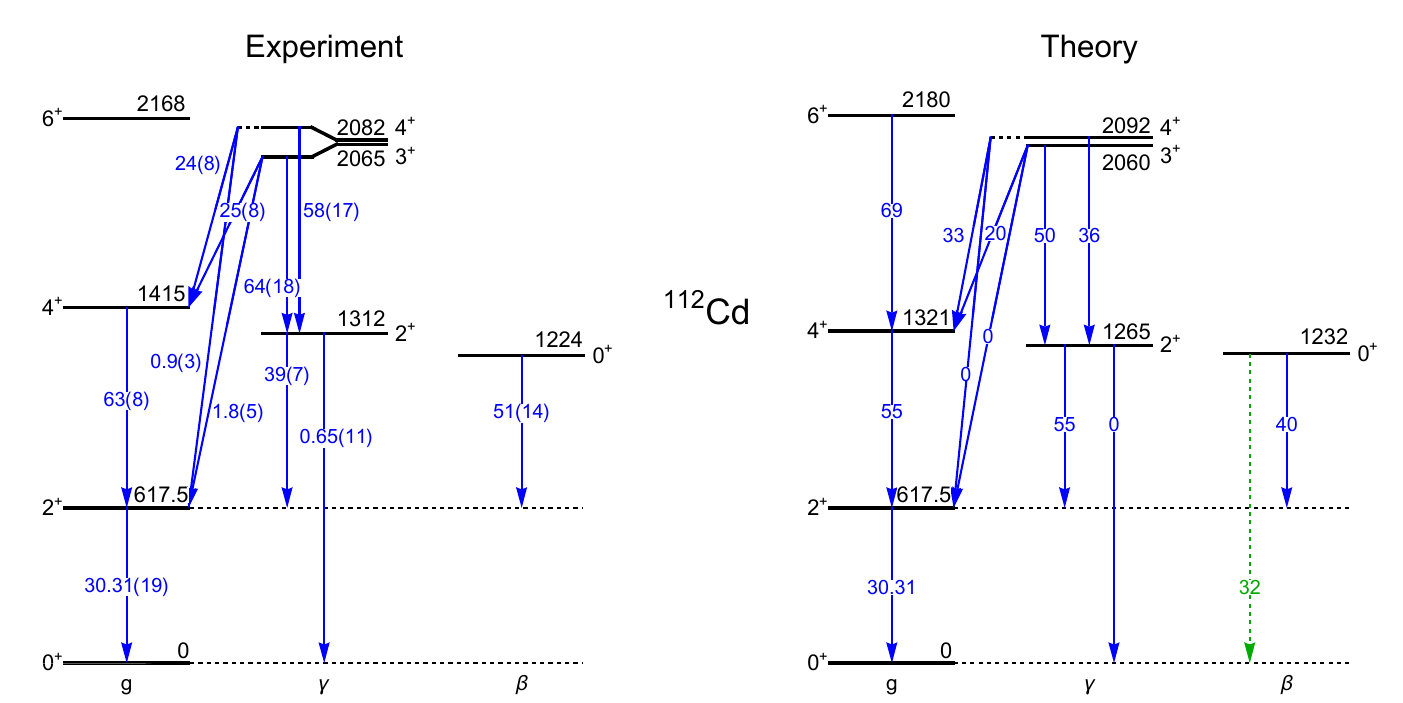} \\
	\caption{(Color online) The same as in Fig. \ref{fig9}, but for the experimental data of $^{112}$Cd \cite{Lalkovski}.}
	\label{fig15}
\end{figure*}

\begin{figure*}
	\centering
\includegraphics[width=.65\textwidth]{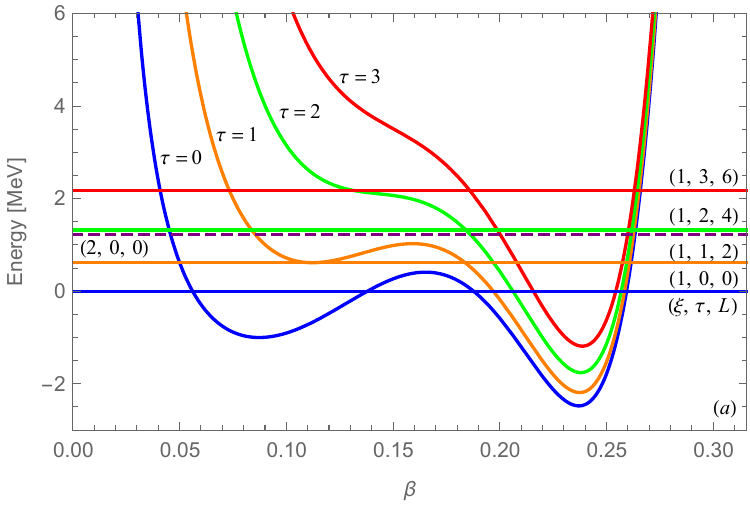}\\
\includegraphics[width=.65\textwidth]{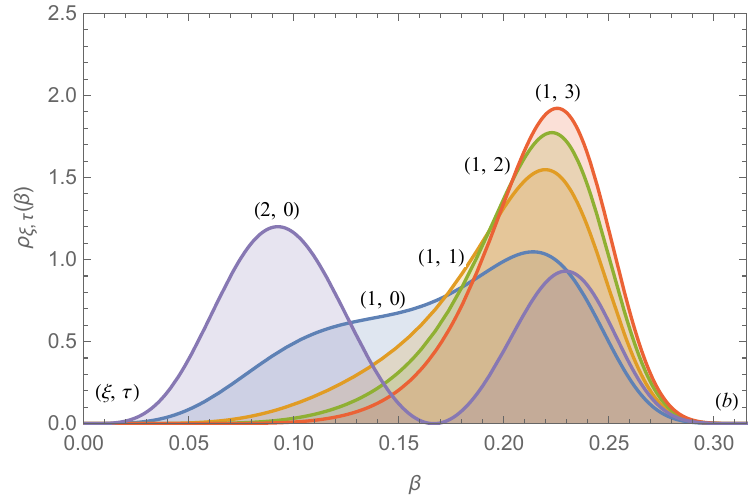}
	\caption{(Color online) The same as in Fig. \ref{fig10}, but for $^{112}$Cd \cite{Lalkovski}.}
	\label{fig16}
\end{figure*}

\begin{figure*}
	\centering
		\includegraphics[width=1.\textwidth]{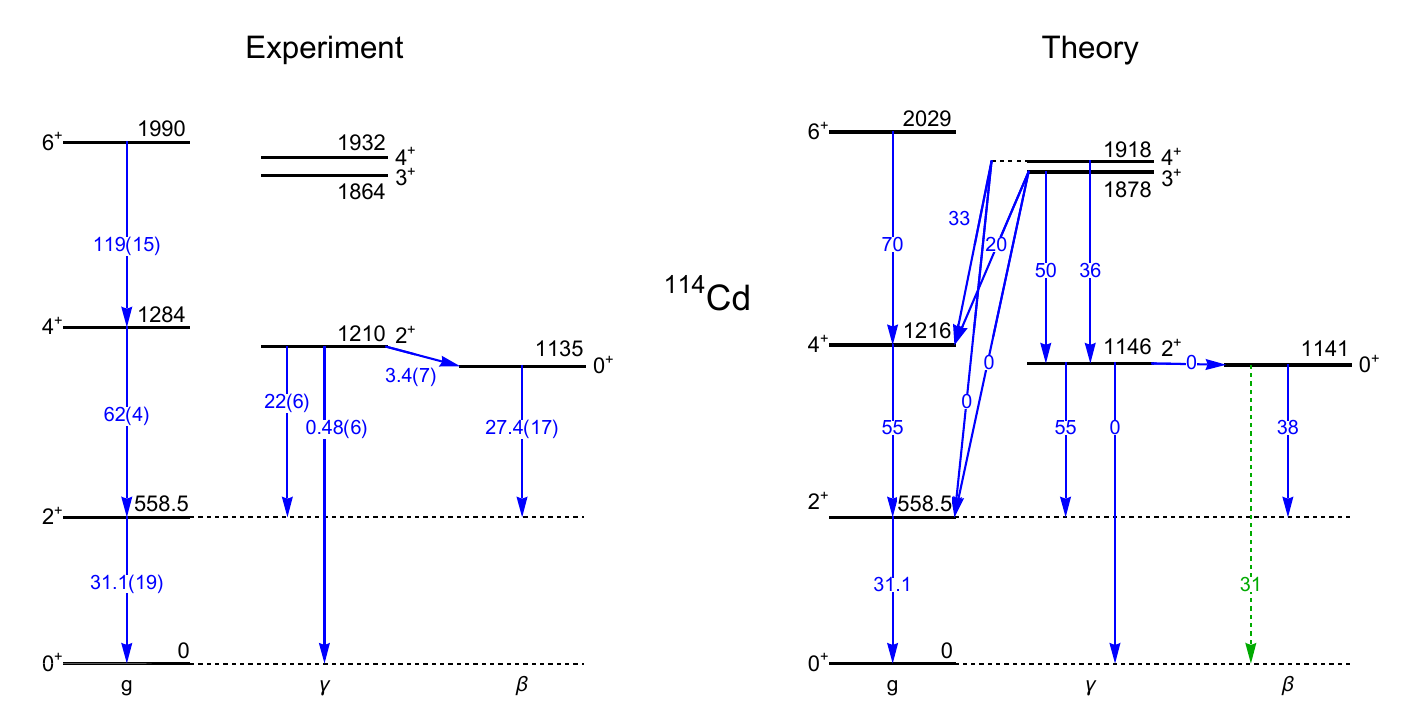} \\
	\caption{(Color online) The same as in Fig. \ref{fig9}, but for the experimental data of $^{114}$Cd \cite{Blachot2}.}
	\label{fig17}
\end{figure*}

\begin{figure*}
	\centering
\includegraphics[width=.65\textwidth]{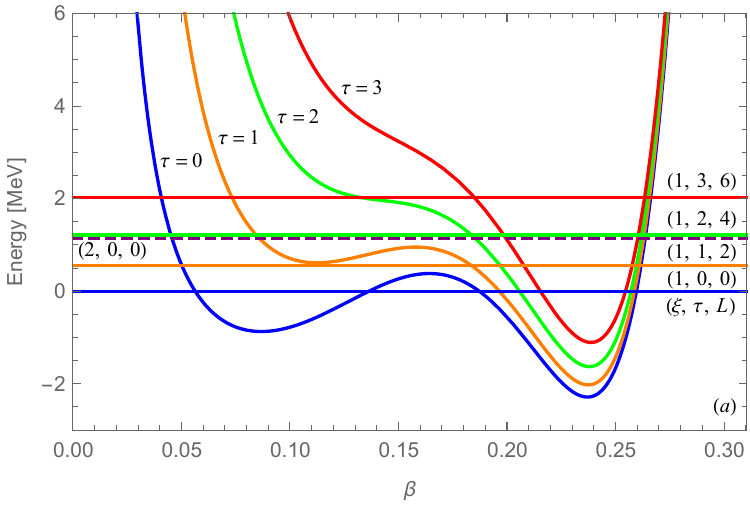}\\
\includegraphics[width=.65\textwidth]{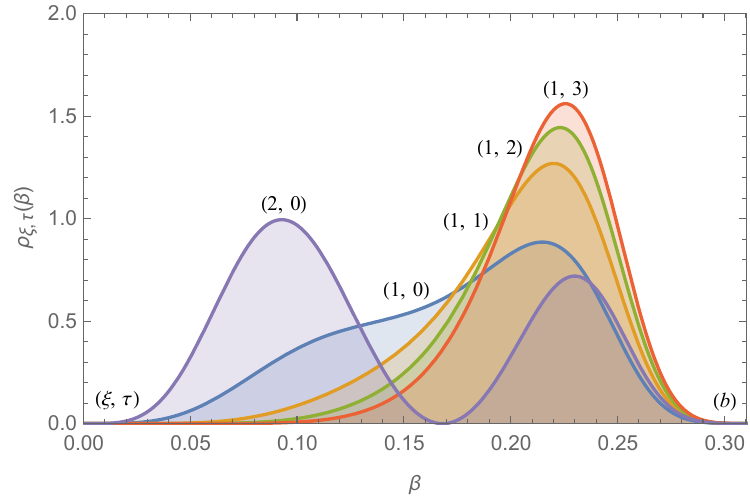}
	\caption{(Color online) The same as in Fig. \ref{fig10}, but for $^{114}$Cd \cite{Blachot2}.}
	\label{fig18}
\end{figure*}

\begin{figure*}
	\centering
		\includegraphics[width=1.\textwidth]{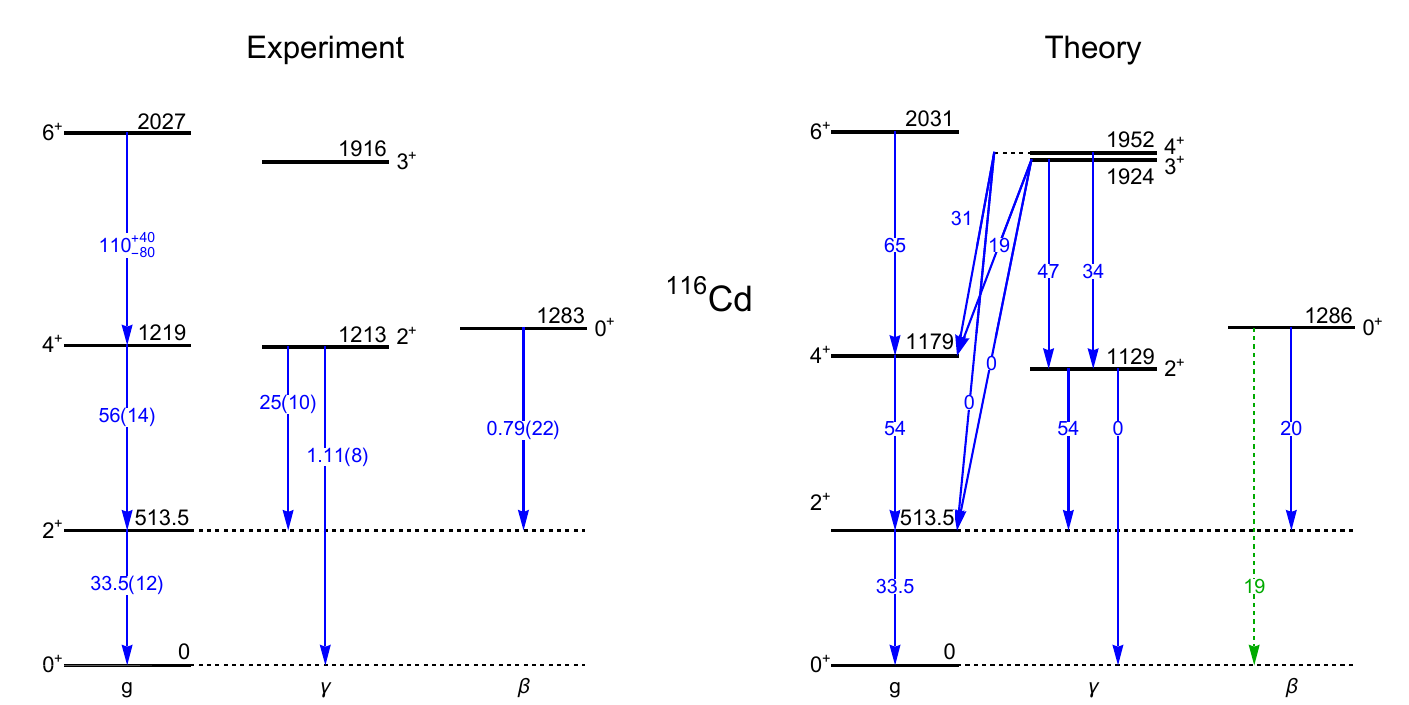} \\
	\caption{(Color online) The same as in Fig. \ref{fig9}, but for the experimental data of $^{116}$Cd \cite{Blachot3}.}
	\label{fig19}
\end{figure*}

\begin{figure*}
	\centering
\includegraphics[width=.65\textwidth]{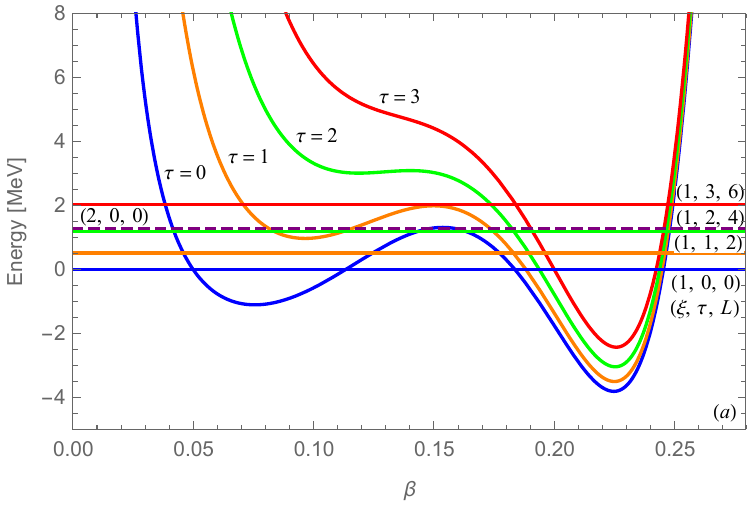}\\
\includegraphics[width=.65\textwidth]{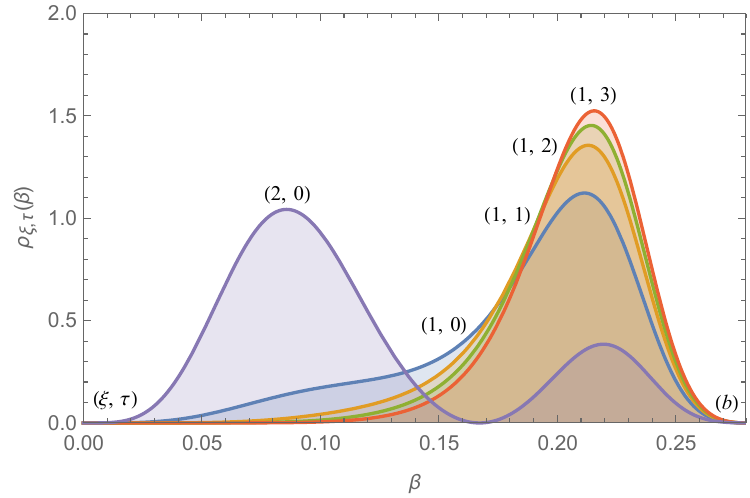}
	\caption{(Color online) The same as in Fig. \ref{fig10}, but for $^{116}$Cd \cite{Blachot3}.}
	\label{fig20}
\end{figure*}

It is well known that the shape coexistence appears in a narrow range of energy for the lowest states, respectively that the shape phase transition is discussed in relation to the change in shape of the ground state along an isotopic chain. Therefore, in the present study are considered only the $0_{g}^{+}$, $2_{g}^{+}$, $4_{g}^{+}$ and $6_{g}^{+}$ states of the ground band, the first excited $0_{\beta}^{+}$ state, respectively the $2_{\gamma}^{+}$, $3_{\gamma}^{+}$ and $4_{\gamma}^{+}$ states of the first $\gamma$-band of the $^{106-116}$Cd isotopes \cite{Frenne,Blachot1,Gurdal,Lalkovski,Blachot2,Blachot3}. This selection increases the ability of the model to see the presence of these phenomena because otherwise, by including higher states in the fit, one can induce some small uncertainties in determination of the properties of these lowest states which are more relevant for the present study. The parameters of the model, given in Table \ref{tab1}, have been determined in the following way. The independent parameters $b_{2}$, $b_{3}$, $b_{4}$ and $\tilde{b}$ are fitted for the experimental energies of the considered states using the root mean square procedure (rms) for the normalized energy formula (\ref{norunst}). The scaling parameter $\hbar^{2}/(2\tilde{B})$, is fixed such that to exactly reproduce the experimental energy of the first excited $2_{g}^{+}$ state of the ground band. Concerning $\tilde{\beta_{w}}$, which defines the infinite square well potential, is set such that to intersect tangentially the tail of the octic potential. This constraint involves a dependence of $\tilde{\beta}_{w}$ on the parameters of the octic potential. Finally, the scaling parameter $\beta_{M}$ is fixed for the $B(E2)$ transition between the first excited $2_{g}^{+}$ state of the ground band and the ground state $0_{g}^{+}$ and it is further employed also to calculate the monopole transition $\rho^{2}(E0)$ (\ref{monopole}). It is also used to cast the model into the space of the $\beta=\beta_{M}\tilde{\beta}$ deformation variable with a traditional range of values. The calculated energies and electromagnetic transitions, as well as the corresponding energy potentials and probability density distributions of deformation, are shown for each isotope in Figs. \ref{fig9}-\ref{fig20}. In the following, a detailed discussion is made on these results.

The energies of the considered states are very well reproduced for all isotopes. An exception comes from energy of the $2_{\gamma}^{+}$ state, which is overestimated initially for $^{106}$Cd and $^{108}$Cd, while in heavier nuclei the agreement is gradually improved. This is reflected also in the rms values listed in Table \ref{tab1} and it can be explained by the fact that in the picture of the $\gamma$-unstable symmetry this state is degenerated with the $4_{g}^{+}$. In the present model the term $L(L+1)$ is added to break this degeneracy, but not to recover such large difference in energy.  On the contrary, the grouping of the $3_{\gamma}^{+}$ and $4_{\gamma}^{+}$, which corresponds to the staggering of the $\gamma$-band of the $\gamma$-unstable system \cite{McCutchan}, is described remarkably well for all isotopes even for the $^{106,108}$Cd isotopes. Another interesting behavior here is the type of variation in the relative energy between the $0_{\beta}^{+}$ and $2_{\gamma}^{+}$ states. Starting with $^{106}$Cd, the $0_{\beta}^{+}$ is higher in energy than the $2_{\gamma}^{+}$. The two states become almost degenerated for $^{110}$Cd, from where $0_{\beta}^{+}$ goes below $2_{\gamma}^{+}$ up to $^{116}$Cd where it goes back above. This trend is followed also in theory, excepting the approximate degeneracy point in $^{110}$Cd, which in theory it starts to manifest to the next isotope $^{112}$Cd and more evident in $^{114}$Cd.

Concerning the electromagnetic transitions, the scaling parameter $\beta_{M}$ is fixed for the experimental value of the $B(E2;2_{g}^{+}\rightarrow0_{g}^{+})$ transition, thus the value for this transition is the same in theory. The next transition in the ground band, namely $B(E2;4_{g}^{+}\rightarrow2_{g}^{+})$ is reproduced with high accuracy for all isotopes. The theoretical value for the $B(E2;6_{g}^{+}\rightarrow4_{g}^{+})$ transition falls within the range of the experimental errors for $^{106}$Cd, $^{110}$Cd and $^{116}$Cd, respectively is somewhat smaller for $^{114}$Cd. Instead, this experimental value is missing for $^{108}$Cd and $^{112}$Cd where the calculated values are given to support future experiments. Other common experimental data are $B(E2;2_{\gamma}^{+}\rightarrow2_{g}^{+})$ and $B(E2;2_{\gamma}^{+}\rightarrow0_{g}^{+})$. The first one is approximatively well described especially for $^{110}$Cd and $^{112}$Cd, while the last one it is very small in magnitude, but clearly it is not forbidden as in theory.

Of great interest is also the transition $B(E2;0_{\beta}^{+}\rightarrow2_{g}^{+})$, experimentally available only for the last three isotopes. The agreement is good here between theory and experiment for $^{112}$Cd and $^{114}$Cd, but not for $^{116}$Cd for which this transition drastically decreases in experiment relative to the values of the previous isotopes. A smoother decline in intensity is obtained also in theory, supporting thus the observed change, which can be ascribed to the sudden increase in the energy of the $0_{\beta}^{+}$ state. It is worth noting here that the value for this transition is very big for $^{112}$Cd and $^{114}$Cd, namely $51_{-14}^{+14}$ [W.u.] (40 [W.u.] in Th.) and $27.4_{-1.7}^{+1.7}$ [W.u.] (38 [W.u.] in Th.). These values are comparable with those from within the ground band and indicates the presence of the shape coexistence with mixing. Indeed, in Figs. \ref{fig16} and \ref{fig18}, for these two isotopes one gets two clear peaks for the probability density distribution of deformation of the ground states even if one has no-node for the wave function. Also, the for the $0_{\beta}^{+}$ the plots look like mirror reflections of the ground state distributions. Instead, in Fig. \ref{fig20}, the mixing of the two states $0_{\beta}^{+}$ and $0_{g}^{+}$ decreases. This results from the fact that for both states one has a more dominant peak, above the well-deformed minimum for the ground state, respectively to the less-deformed minimum for the $0_{\beta}^{+}$. The decreasing of mixing explains here the decreasing in the $B(E2)$ value. A lower experimental value for this transition could be achieved then by inhibiting the mixing between $0^{+}$ states. A signature for the presence of the coexistence with mixing is given by the monopole transition between these two states (\ref{monopole}). There are no experimental data for these three isotopes, but the theoretical values indicated by dashed arrows in Figs. \ref{fig15}, \ref{fig17} and \ref{fig19} support a presence of the shape coexistence with mixing for these three isotopes.

Other available experimental $B(E2)$ transitions are for $^{110}$Cd and especially for $^{112}$Cd where the presence of a shape coexistence with mixing is observed. For example, in the case of $^{112}$Cd in Fig. \ref{fig15}, one has additional $E2$ transitions in $\gamma$ band and from the $\gamma$ band to the ground band: $B(E2;4_{\gamma}^{+}\rightarrow2_{\gamma}^{+})=58(17)$ [W.u.] (36 [W.u.] in Th.), $B(E2;4_{\gamma}^{+}\rightarrow4_{g}^{+})=24(8)$ [W.u.] (33 [W.u.] in Th.), $B(E2;4_{\gamma}^{+}\rightarrow2_{}^{+})=0.9(3)$ [W.u.] (forbidden in Th.), $B(E2;3_{\gamma}^{+}\rightarrow2_{\gamma}^{+})=64(18)$ [W.u.] (50 [W.u.] in Th.), $B(E2;3_{\gamma}^{+}\rightarrow4_{g}^{+})=25(8)$ [W.u.] (20 [W.u.] in Th.), $B(E2;3_{\gamma}^{+}\rightarrow2_{g}^{+})=1.8(5)$ [W.u.] (forbidden in Th.). The agreement between theory and experiment is very good also for these transitions if the experimental errors are taken in consideration, excepting the two forbidden ones which are indeed small but non-zero in experiment.

After the previous analysis, one can discuss about shape coexistence in $^{112,114,116}$Cd. For $^{110}$Cd, in Fig. \ref{fig14}, a potential with two minima is obtained. However. the two minima do not have a very deep, such that the states drop below the maximum (barrier). Consequently, the plot of the probability density distribution of deformation indicates a shape fluctuation for the ground state and a regular $\beta$-vibration mode for the $0_{\beta}^{+}$. With other words, there is no splitting of the extended peak of the ground state or clear dominant peak for one minimum of the potential for the $0_{\beta}^{+}$ state. This picture corresponds more to the critical point of the shape phase transition. Indeed, for the lighter isotope $^{108}$Cd one has a flat potential which is specific to the critical point of a second order shape phase transition from an approximately spherical shape to a $\gamma$-unstable deformation \cite{Iachello1}. It is interesting that this isotope chain of Cadmium was investigated in \cite{Lahbas}, using the quasi-exact solution of the Bohr-Mottelson Hamiltonian with a sextic oscillator potential, getting that $^{108}$Cd is the best candidate for the critical point of this shape phase transition. Going back to $^{106}$Cd, one can see in Fig. \ref{fig10} a potential with a more flat minimum and a clear deformed one. The tail of the probability density distribution of deformation of the ground state is extended also on the first less-deformed minimum, while the excited states of the ground band have a single peak centered around the well-deformed minimum. For the first excited $0^{+}$ one has a regular distribution of deformation. Correlating this picture with the calculated small monopole $E0$ transition of $16$ units, given in Fig. \ref{fig9} by dashed arrow, one can conclude that this isotope fits in the category of a $\gamma$-unstable deformation with shape fluctuations in the ground state. These results for $^{106-116}$Cd are very promising from the perspective of the future applications of the model proving thus its ability to offer a complete description of the lowest states.

\begin{table}
\caption{The model parameters determined for the experimental data of the $^{106-116}$Cd isotope chain \cite{Frenne,Blachot1,Gurdal,Lalkovski,Blachot2,Blachot3}. Starting with the lighter isotope, the corresponding parameters lead to the following values for $\tilde{\beta}_w$: 1.93, 2.00, 2.30, 2.30, 2.79, 3.09.}
\begin{center}
\resizebox{10cm}{1.5cm} {
\begin{tabular}{ccccccccc}
\hline
\hline
Nucleus&$\frac{\hbar}{2\tilde{B}}$ [keV]&$b_{2}$&$b_{3}$&$b_{4}$&$\tilde{b}$&$\beta_{M}$&$rms$\\
\hline
$^{106}$Cd&163&36.408&$-$38.631&9.899&0.038&0.146&0.236\\
$^{108}$Cd&107&$-$4.122&$-$0.192&1.864&0.106&0.178&0.192\\
$^{110}$Cd&203&5.885&$-$3.998&0.673&0.049&0.132&0.103\\
$^{112}$Cd&215&22.298&$-$15.242&2.573&0.019&0.137&0.071\\
$^{114}$Cd&300&5.835&$-$2.702&0.302&0.017&0.111&0.076\\
$^{116}$Cd&475&3.381&$-$1.197&0.100&0.008&0.090&0.081\\
\hline
\hline
\end{tabular}}
\end{center}
\label{tab1}
\end{table}







\section{Conclusions}
The Bohr-Mottelson Hamiltonian with an octic oscillator potential in the $\beta$ variable and $\gamma$-unstable symmetry is numerically solved in a basis of functions which are the solutions for an infinite square well potential. The basis functions are expressed in terms of the Bessel functions of the first kind. Numerical calculations evidence the ability of the model to describe phase transition and its critical point with respect to the shape of the ground state, shape transition in band as a function of the total angular momentum (dynamical phase transitions), shape coexistence with and without mixing, respectively associated quantities as unusually small intra-band $(E2)$ transitions, large inter-band $(E2)$ transitions or large monopole transition $(E0)$ between the first excited $0_{2}^+$ state and the ground state $0_{1}^+$. These remarks are supported by previous applications of the sextic potential to many experimental data, the sextic potential being a particular case of the octic potential, but also by the present application of the model to the $^{106-116}$Cd isotopes.
The octic potential is obviously more versatile than the sextic one. One particular feature pertains to the possibility to have a wider separating barrier between distinct deformation minima. This is essential for a more nuanced description of the mixing between coexisting deformation configurations. This aspect was more evident in the investigation of the isotopes of Cadmium, where a more accurate description of the experimental data could be obtained. These results indicate the presence of a shape coexistence with mixing in $^{112,114,116}$Cd, specific shape fluctuations of the critical point in $^{108,110}$Cd, respectively a pronounced $\gamma$-unstable deformation with shape fluctuations in the ground state for $^{106}$Cd.

It is worth mentioning here that this numerical diagonalization method for the octic potential, which will be definetly extended to other nuclear deformation conditions, such as prolate and stable-triaxial ones, can also be adapted to physics problems in other fields  where one needs to solve the Schr\"{o}dinger equation for a polynomial potential. Thus, the general scholarly character of the present model, extends is applicability beyond the field of nuclear physics.

\section{Acknowledgments}
This work was supported by grants of the Ministry of Research, Innovation and Digitization, CNCS-UEFISCDI, project number PN-IV-P1-PCE-2023-0273, within PNCDI IV, and project number PN-23-21-01-01/2023.


\begin{thebibliography}{00}


\bibitem{Bohr1} A. Bohr, Mat. Fys. Medd. Dan. Vid. Selsk. 26 (1952) No. 14.
\bibitem{Bohr2} A. Bohr and B. R. Mottelson, Mat. Fys. Medd. Dan. Vid. Selsk. 27 (1953) No. 16.
\bibitem{Fortunato1} L. Fortunato, Eur. Phys. J. A 26 (2005) 1.
\bibitem{Casten} R. F. Casten, Nat. Phys. 2 (2006) 811.
\bibitem{Cejnar} P. Cejnar, J. Jolie and R. F. Casten, Rev. Mod. Phys. 82 (2010) 2155.
\bibitem{Buganu1} P. Buganu and L. Fortunato, J. Phys. G: Nucl. Part. Phys. 43 (2016) 093003.
\bibitem{Fortunato2} L. Fortunato, Prog. Part. Nucl. Phys. 121 (2021) 103891.
\bibitem{Morinaga} H. Morinaga, Phys. Rev. 101 (1956) 254.
\bibitem{Heyde} K. Heyde and J. L. Wood, Rev. Mod. Phys. 83 (2011) 1467.
\bibitem{Martinou} A. Martinou, D. Bonatsos, T. J. Mertzimekis, K. E. Karakatsanis, I. E. Assimakis, S. K. Peroulis, S. Sarantopoulou and N. Minkov, Eur. Phys. J. A 57 (2021) 84.
\bibitem{Garrett} P. E. Garrett, M. Zieli\'{n}ska and E. Cl\'{e}ment, Prog. Part. Nucl. Phys. 124 (2022) 103931.
\bibitem{Bonatsos} D. Bontasos, A. Martinou, S. K. Peroulis, T. J. Mertzimekis and N. Minkov, J. Phys. G: Nucl. Part. Phys. 50 (2023) 075105.
\bibitem{Bona} D. Bontasos, A. Martinou, S. K. Peroulis, T. J. Mertzimekis and N. Minkov, Atoms 11 (2023) 117.
\bibitem{Mayer} M. G. Mayer and J. H. D. Jensen, {\it Elementary Theory of Nuclear Shell Structure}, (Wiley: New York, NY, USA, 1955).
\bibitem{Poves1} A. Poves, J. Phys. G: Nucl. Part. Phys. 43 (2016) 024010.
\bibitem{Iachello} F. Iachello and A. Arima, {\it The Interacting Boson Model}, (Cambridge U. Press, Cambridge, UK., 1987).
\bibitem{Nomura} K. Nomura, T. Otsuka and P. van Isacker, J. Phys. G: Nucl. Part. Phys. 43 (2016) 024008.
\bibitem{Maya} E. Maya-Barbecho, S. Baid, J. M. Arias and J. E. Garc\'{i}a-Ramos, Phys. Rev. C 108 (2023) 034316.
\bibitem{Cheng} Cheng-Fu Mu and Da-Li Zhang, Commun. Theor. Phys. 74 (2022) 025302.
\bibitem{Leviatan} A. Leviatan and N. Gavrielov, Phys. Scr. 92 (2017) 114005.
\bibitem{Martinou1} A. Martinou, D. Bonatsos, S. K. Peroulis, K. E. Karakatsanis, T. J. Mertzimekis and N. Minkov, Symmetry 15 (2023) 29.
\bibitem{Martinou2} D. Bonatsos, A. Martinou, S. K. Peroulis, T. J. Mertzimekis and N. Minkov, Symmetry 15 (2023) 169.
\bibitem{Yang} Y. L. Yang, P. W. Zhao and Z. P. Li, Phys. Rev. C 107 (2023) 024308.
\bibitem{Niksic} T. Nik\v{s}i\'{c}, D. Vretenar and P. Ring, Phys. Rev. C 78 (2008) 034318.
\bibitem{Zhou} E. F. Zhou, C. R. Ding, J. M. Yao, B. Bally, H. Hergert, C. F. Jiao and T. R. Rodr\'{i}guez, Phys. Lett. B 865 (2025) 139464.
\bibitem{Hu} B. S. Hu, Z. H. Sun, G. Hagen and T. Papenbrock, Phys. Rev. C 110 (2024) L011302.
\bibitem{Majola} S. N. T. Majola and \textit{et al.}, Phys. Rev. C 100 (2019) 044324.
\bibitem{Matsuyanagi} K. Matsuyanagi, M. Matsuo, T. Nakatsukasa, K. Yoshida, N. Hinohara and K. Sato, J. Phys. G: Nucl. Part. Phys. 43 (2016) 024006.
\bibitem{Mennana1} A. Ait Ben Mennana, R. Benjedi, R. Budaca, P. Buganu, Y. El Bassem, A. Lahbas and M. Oulne, Phys. Scr. 96 (2021) 125306.
\bibitem{Mennana2} A. Ait Ben Mennana, R. Benjedi, R. Budaca, P. Buganu, Y. El Bassem, A. Lahbas and M. Oulne, Phys. Rev. C 105 (2022) 034347.
\bibitem{Benjedi} R. Benjedi, R. Budaca, P. Buganu, Y. El Bassem, A. Lahbas and M. Oulne, Phys. Scr. 99 (2022) 055307.
\bibitem{Buganu2} P. Buganu, S. Chafik, A. Lahbas and M. Oulne, Symmetry 17 (2025) 687.
\bibitem{Iachello1} F. Iachello, Phys. Rev. Lett. 85 (2000) 3580.
\bibitem{Iachello2} F. Iachello, Phys. Rev. Lett. 87 (2001) 052502.
\bibitem{Ushveridze} A. G. Ushveridze, {\it Quasi-exactly Solvable Models in Quantum Mechanics}, (IOP, Bristol, 1994).
\bibitem{Budaca} R. Budaca, P. Buganu, M. Chabab, A. Lahbas and M. Oulne, Ann. Phys. (NY) 375 (2016) 65.
\bibitem{Levai} G. L\'{e}vai and J. M. Arias, Symmetry 15 (2023) 2059.
\bibitem{Levai2} G. L\'{e}vai and J. M. Arias, Phys. Rev. C 69 (2004) 014304.
\bibitem{Levai3} G. L\'{e}vai and J. M. Arias, Phys. Rev. C 81 (2010) 044304.
\bibitem{Gneuss} G. Gneuss, U. Mosel and W. Greiner, Phys. Lett. B 30 (1969) 397.
\bibitem{Gneuss2} G. Gneuss and W. Greiner, Nucl. Phys. A 171 (1971) 449.
\bibitem{Hess} P. O. Hess, M. Seiwert, J. Maruhn and W. Greiner, Z. Phys. A 296 (1980) 147.
\bibitem{Eisenberg} J. M. Eisenberg and W. Greiner, {\it Nuclear Models: Collective and Single-Particle Phenomena}, Nuclear Theory, Vol. 1, (Publisher North-Holland, 1987).
\bibitem{ACM1} D. J. Rowe and J. L. Wood, {\it Fundamentals of Nuclear Models: Foundational Models}, (World Scientific, Singapore, 2010).
\bibitem{ACM2} T. A. Welsh and D. J. Rowe, Comput. Phys. Commun. 200 (2016) 220.
\bibitem{Budaca1} R. Budaca, P. Buganu and A. I. Budaca, Phys. Lett. B 776 (2018) 26.
\bibitem{Budaca2} R. Budaca, A. I. Budaca and P. Buganu, J. Phys. G: Nucl. Part. Phys. 46 (2019) 125102.
\bibitem{BudacaA} R. Budaca and A. I. Budaca, EPL 123 (2018) 42001.
\bibitem{Budaca3} R. Budaca, P. Buganu and A. I. Budaca, Nucl. Phys. A 990 (2019) 137.
\bibitem{Buganu3} P. Buganu, R. Benjedi and M. Oulne, Mathematics 13 (2025) 460.
\bibitem{Bes} D. R. B\`{e}s, Nucl. Phys. 10 (1959) 373.
\bibitem{Bonatsos1} D. Bonatsos, D. Lenis, D. Petrellis and P. A. Terziev, Phys. Lett. B 588 (2004) 172.
\bibitem{Bonatsos2} D. Bonatsos, D. Lenis, D. Petrellis, P. A. Terziev and I. Yigitoglu, Phys. Lett. B 621 (2005) 102.
\bibitem{Bonatsos3} D. Bonatsos, D. Lenis, D. Petrellis, P. A. Terziev and I. Yigitoglu, Phys. Lett. B 632 (2006) 238.
\bibitem{Taseli} H. Ta\c{s}eli and A. Zafer, Int. J. Quant. Chem. 63 (1997) 935.
\bibitem{Wilets} L. Wilets and M. Jean, Phys. Rev. 102 (1956) 788.
\bibitem{Caprio} M. A. Caprio and F. Iachello, Nucl. Phys. A 781 (2007) 26.
\bibitem{Wigner} E. P. Wigner, {\it Group Theory and its Application to the Quantum Mechanics of Atomic Spectra}, (Ed. J J Griffin (Academic Press Inc., 1959).
\bibitem{Rowe1} D. J. Rowe, P. S. Turner and J. Repka, J. Math. Phys. 45 (2004) 2761.
\bibitem{Rowe2} D. J. Rowe, J. Phys. A: Math. Gen. 38 (2005) 10181.
\bibitem{Rowe3} D. J. Rowe and P. S. Turner, Nucl. Phys. A 753 (2005) 94.
\bibitem{Wood} J. L. Wood, E. F. Zganjar, C. de Coster and K. Heyde, Nucl. Phys. A 651 (1999) 323.
\bibitem{Ginocchio} J. H. Ginocchio and M. W. Kirson, Phys. Rev. Lett. 44 (1980) 1744.
\bibitem{Dieperink} A. E. L. Dieperink, O. Scholten and F. Iachello, Phys. Rev. Lett. 44 (1980) 1747.
\bibitem{Lahbas} A. Lahbas, P. Buganu and R. Budaca, Mod. Phys. Lett. A 35 (2020) 2050085.
\bibitem{Kumar} P. Kumar and S. K. Dhiman, Nucl. Phys. A 1001 (2020) 121935.
\bibitem{Garrett2} P. E. Garrett and \emph{et al}., Phys. Rev. Lett. 123 (2019) 142502.
\bibitem{Nomura2} K. Nomura and K. E. Karakatsanis, Phys. Rev. C 106 (2022) 064317.
\bibitem{Sharma} S. Sharma, R. Devi and S. K. Khosa, Int. J. Mod. Phys. E 31 (2022) 2250104.
\bibitem{Bohr3} A. Bohr and B. R. Mottelson, {\it Nuclear Structure}, Vol. II, (World Scientific, Singapore, 1998).
\bibitem{Bernard} B. L. Cohen and R. E. Price, Phys. Rev. 118 (1960) 1582.
\bibitem{Peker} R. Meyer and L. Peker, Z. Phys. A 283 (1977) 379.
\bibitem{Heyde2} K. Heyde, P. Van Isacker, M. Waroquier and G. Wenes, Phys. Rev. C 25 (1982) 3160.
\bibitem{Casten2} R. F. Casten, {\it et al.}, Phys. Lett. B 297 (1992) 19.
\bibitem{Wrzosek}  K. Wrzosek-Lipska, {\it et al.}, Acta Phys. Pol. B 51 (2020) 789.
\bibitem{Leviatan2} A. Leviatan, N. Gavrielov, J. E. Garc\'{i}a-Ramos and P. Van Isacker, Phys. Rev. C 98 (2018) 031302(R).
\bibitem{Gavrielov}  N. Gavrielov, J. E. Garc\'{i}a-Ramos, P. Van Isacker and A. Leviatan, Phys. Rev. C 108 (2023) L031305.
\bibitem{Wang} T. Wang, X. Chen and Y. Zhang, Chin. Phys. C 49 (2025) 014107.
\bibitem{Garrett3} P. E. Garrett, J. Phys. G: Nucl. Part. Phys. 27 (2001) R1.
\bibitem{Garrett4} P. E. Garrett, J. L. Wood and S. W. Yates, Phys. Scr. 93 (2018) 063001.
\bibitem{Sharpey} J. F. Sharpey-Schafer, R. A. Bark, S. P. Bvumbi, T. R. S. Dinoko and S. N. T. Majola, Eur. Phys. J. A 55 (2019) 15.
\bibitem{Garrett5} P. E. Garrett, K. L. Green and J. L. Wood, Phys. Rev. C 78 (2008) 044307.
\bibitem{Frenne} D. De Frenne and A. Negret, Nucl. Data Sheets 109 (2008) 943.
\bibitem{Blachot1} J. Blachot, ENSDF (2008).
\bibitem{Gurdal} G. G$\ddot{u}$rdal and F. G. Kondev, Nucl. Data Sheets 113 (2012) 1315.
\bibitem{Lalkovski} S. Lalkovski and F. G. Kondev, Nucl. Data Sheets 124 (2015) 157.
\bibitem{Blachot2} J. Blachot, Nucl. Data Sheets 113 (2012) 515.
\bibitem{Blachot3} J. Blachot, Nucl. Data Sheets 111 (2010) 717.
\bibitem{McCutchan} E. A. McCutchan, D. Bonatsos, N. V. Zamfir and R. F. Casten, Phys. Rev. C 76 (2007) 024306.
\end{thebibliography}



\end{document}